\documentclass[twocolumn,aps,prx,superscriptaddress, 10pt]{revtex4-2}

\usepackage[english]{babel}
\usepackage{graphicx}
\usepackage{dcolumn}
\usepackage[dvipsnames]{xcolor}
\usepackage[mathlines]{lineno}
\usepackage{braket}
\usepackage{tabularx}
\usepackage{amsmath}
\usepackage{mathtools}
\usepackage[noend]{algpseudocode}
\usepackage{algorithm}
\usepackage{hyperref}
\usepackage{amsfonts}
\usepackage{amssymb}
\usepackage{dsfont}
\usepackage{subfigure}
\usepackage{subcaption}
\usepackage{csquotes}
\usepackage{ragged2e}
\usepackage{booktabs}
\usepackage[normalem]{ulem}

\newcommand{\mus}[0]{{\, \mu \text{s}}}
\newcommand{\MHz}[0]{{\, \text{MHz}}}
\newcommand{\kHz}[0]{{\, \text{kHz}}}

\newcommand{\vtheta}[0]{\boldsymbol{\theta}}

\newcommand{\vA}[0]{{\boldsymbol{A}}}
\newcommand{\vx}[0]{{\boldsymbol{x}}}
\newcommand{\vy}[0]{{\boldsymbol{y}}}
\newcommand{\vz}[0]{{\boldsymbol{z}}}

\newcommand{\vomega}[0]{\boldsymbol{\omega}}
\newcommand{\vmu}[0]{{\boldsymbol{\mu}}}

\newcommand{\vphi}[0]{{\boldsymbol{\varphi}}}
\newcommand{\vlambda}[0]{{\boldsymbol{\lambda}}}

\newcommand{\var}[1]{{\text{Var}(#1)}}

\newcommand{\norm}[1]{\left\lVert#1\right\rVert}

\newcommand{\class}[1]{\mathfrak{#1}}

\begin{document}
	
\title{A multi-dimensional quantum estimation and model learning framework \\ based on variational Bayesian inference}

\author{Federico Belliardo}
\email {federico.belliardo@gmail.com}
\affiliation{ 
    Institute of Photonics and Quantum Sciences, SUPA, School of Engineering and Physical Sciences, Heriot-Watt University, Edinburgh EH14 4AS, UK
}
\affiliation{ 
    Pritzker School of Molecular Engineering, University of Chicago, Chicago, Illinois 60637, USA
}

\author{Erik M. Gauger}
\affiliation{ 
    Institute of Photonics and Quantum Sciences, SUPA, School of Engineering and Physical Sciences, Heriot-Watt University, Edinburgh EH14 4AS, UK
}

\author{Mohamed H. Abobeih}
\affiliation{ 
QuTech, Delft University of Technology, PO Box 5046, 2600 GA Delft, The Netherlands}
\affiliation{Kavli Institute of Nanoscience Delft, Delft University of Technology, PO Box 5046, 2600 GA Delft, The Netherlands}

\author{Tim H. Taminiau}
\affiliation{ 
QuTech, Delft University of Technology, PO Box 5046, 2600 GA Delft, The Netherlands}
\affiliation{Kavli Institute of Nanoscience Delft, Delft University of Technology, PO Box 5046, 2600 GA Delft, The Netherlands}

\author{Yoann Altmann}
\email {y.altmann@hw.ac.uk}
\affiliation{ 
    Intitute of Signals, Sensors and Systems, School of Engineering and Physical Sciences,
    Heriot-Watt University, Edinburgh EH14 4AS, UK
}

\author{Cristian Bonato}
\email {c.bonato@hw.ac.uk}
\affiliation{ 
    Institute of Photonics and Quantum Sciences, SUPA, School of Engineering and Physical Sciences, Heriot-Watt University, Edinburgh EH14 4AS, UK
}

\begin{abstract}
The advancement and scaling of quantum technology has made the learning and identification of quantum systems and devices in highly-multidimensional parameter spaces a pressing task for a variety of applications. In many cases, the integration of real-time feedback control and adaptive choice of measurement settings places strict demands on the speed of this task. 
    
Here we present a joint model selection and parameter estimation algorithm that is fast and operable on a large number of model parameters. The algorithm is based on variational Bayesian inference (VBI), which approximates the target posterior distribution by optimizing a tractable family of distributions, making it more scalable than exact inference methods relying on sampling and that generally suffer from high variance and computational cost in high-dimensional spaces. We show how a regularizing prior can be used to select between competing models, each comprising a different number of parameters, identifying the simplest model that explains the experimental data. The regularization can further separate the degrees of freedom, e.g.~quantum systems in the environment or processes, which contribute to major features in the observed dynamics, with respect to others featuring small coupling, which only contribute to a background. 
    
As an application of the introduced framework, we consider the problem of the identification of multiple individual nuclear spins with a single electron-spin quantum sensor, relevant for nanoscale nuclear magnetic resonance and for the implementation of multi-qubit quantum networking nodes. With the number of environmental spins unknown a priori, our Bayesian approach is able to correctly identify the model, i.e.~the number of spins and their couplings. We benchmark the algorithm on both simulated and experimental data, using standard figures of merit, and demonstrate that we can estimate dozens of parameters within minutes. Our methodology is compatible with the implementation of real-time adaptive choice of experimental settings in multi-parameter estimation, which has the potential to greatly decrease the estimation time in a variety of metrology, tomography and characterization tasks by prioritizing the measurements with the highest content of information. 
\end{abstract}

\maketitle

\vspace{5mm}

\section{Introduction}
\label{sec:introduction}
The increasingly fast-paced progress in the scale and complexity of quantum devices developed for quantum computing \cite{bravyi_HighthresholdLowoverheadFaulttolerant_2024, googlequantumaiandcollaborators_QuantumErrorCorrection_2025, maskara_ProgrammableSimulationsMolecules_2025, bluvstein_ArchitecturalMechanismsUniversal_2025, chiu_ContinuousOperationCoherent_2025a}, networking \cite{stas_RobustMultiqubitQuantum_2022, hermans_QubitTeleportationNonneighbouring_2022, delledonne_OperatingSystemExecuting_2025,stolk_MetropolitanscaleHeraldedEntanglement_2024, knaut_EntanglementNanophotonicQuantum_2024} and sensing \cite{boto_MovingMagnetoencephalographyRealworld_2018, stray_QuantumSensingGravity_2022, aslam_QuantumSensorsBiomedical_2023, budakianRoadmapNanoscaleMagnetic2024, briegel_OpticalWidefieldNuclear_2025}  {requires} the development of new frameworks to characterize, measure, and `learn' quantum systems {\cite{mohseni_DeepLearningQuantum_2022, gebhartLearningQuantumSystems2023, krenn_ArtificialIntelligenceMachine_2023,dawid_MachineLearningQuantum_2025, acampora_QuantumComputingArtificial_2025,  alexeev_ArtificialIntelligenceQuantum_2025, du_ArtificialIntelligenceRepresenting_2025}}.

Historically, the characterization of quantum systems has broadly followed two approaches in quantum information theory: The first originates in quantum tomography~\cite{blume-kohoutOptimalReliableEstimation2010, parisQuantumStateEstimation2004, grafensteinCoherentSignalDetection2025} and is based on learning the full representation of quantum states and processes, and more recently also their dynamical generators \cite{wiebeQuantumHamiltonianLearning2014,wang_ExperimentalQuantumHamiltonian_2017,anshu_SampleefficientLearningInteracting_2021}, such as the full set of Hamiltonian and/or Lindbladian operators describing the system evolution. In this scenario, the only assumption is that the state to be identified belongs to a large set of physical states. The second approach, rooted in quantum parameter estimation, aims at estimating a typically small set of parameters on which the state depends ~\cite{hayashiAsymptoticTheoryQuantum, holevoProbabilisticStatisticalAspects2011}, with the goal of evaluating the ultimate precision limit on the knowledge of said parameters in terms of the (quantum) Cramér-Rao bound. In this scenario, we assume {that we have} a parametrized expression for the state, or the channel, that we want to characterize, which usually belongs to a small and strongly constrained set of states. The rapidly expanding fields of quantum sensing and metrology are primarily based on quantum parameter estimation, and typically deal with the practical problem of determining the external parameters encoded {in} a quantum system used as a probe, such as {a} magnetic field, temperature, or optical phase. In some cases, the characterization can be extended to model learning scenarios \cite{gentileLearningModelsQuantum2021, wallace_LearningDynamicsMarkovian_2025}, where we want to identify a description of the quantum system among multiple different classes of models, each comprising a different set of free parameters. In this manuscript, we take  an integrated approach  {that bases} the characterization of quantum systems and the optimal selection of the model on parameter estimation.
Although parameter estimation scenarios typically involve fewer parameters than corresponding tomography scenarios, the number of {parameters} could be sufficiently large that the task of model learning is in the regime of high-dimensional parameter estimation, even for relatively simple quantum systems.

The successful characterization of a quantum system should include a  description that is sufficiently accurate for the purposes of the experimenter \cite{gebhartLearningQuantumSystems2023}. We require this description to include statistically significant confidence levels and a clear delimitation of the domain of validity of the performed model selection routine (i.e.~clarity in the set of possible models that have been considered for the learning). The gold standard for satisfying both {requirements} is to produce a Bayesian distribution over the set of models and parameters considered~\cite{lukens_PracticalEfficientApproach_2020}. 

A crucial benefit of Bayesian inference is that, from the probability density or mass function, one can compute uncertainties for every function of the model parameters and optimize experimental settings for subsequent measurements, for example to minimize the overall sensing time. This is very important in quantum applications, which are often limited by slow data acquisition because the signals produced by individual quantum systems — such as streams of single photons or electrons — are generally weak.

Bayesian inference has been widely applied in the context of quantum technology: Bayesian quantum tomography \cite{jonesPrinciplesQuantumInference1991, buzekReconstructionQuantumStates1998, schackQuantumBayesRule2001, kravtsovExperimentalAdaptiveBayesian2013, dimatteo_OperationalGaugefreeQuantum_2020} has been shown to be accurate \cite{blume-kohoutOptimalReliableEstimation2010} and to provide reliable and tight error bounds \cite{christandlReliableQuantumState2012}. Although rarely analytically tractable, numerical approaches have been deployed to make tomography experimentally practical \cite{granadePracticalBayesianTomography2016}. Its efficiency has been further improved with adaptive selection of experimental settings \cite{huszarAdaptiveBayesianQuantum2012, granadePracticalAdaptiveQuantum2017} and adaptive discovery of priors \cite{lukens_PracticalEfficientApproach_2020, mondalBayesianQuantumState2023}. Bayesian techniques can also be used to learn the generators of dynamics, such as the Hamiltonian \cite{wiebeHamiltonianLearningCertification2014, wiebeEfficientBayesianPhase2016, shulmanSuppressingQubitDephasing2014, gentileLearningModelsQuantum2021, gebhart_BayesianQuantumMultiphase_2021, berrittaEfficientQubitCalibration2025} or the Lindblad master equation \cite{wallace_LearningDynamicsMarkovian_2025, fioroniLearningAgentbasedApproach2025}. Applications of these concepts in the context of quantum sensing have demonstrated the adaptive choice of optimal experimental settings to minimize sensing time \cite{bonatoOptimizedQuantumSensing2016, lennonEfficientlyMeasuringQuantum2019, santagatiMagneticFieldLearningUsing2019, valeriExperimentalAdaptiveBayesian2020, dushenkoSequentialBayesianExperiment2020, kaubrueggerQuantumVariationalOptimization2021, joasOnlineAdaptiveQuantum2021, craigie_ResourceefficientAdaptiveBayesian_2021, daurelioExperimentalInvestigationBayesian2022, caouette-mansourRobustSpinRelaxometry2022, zoharRealtimeFrequencyEstimation2023, arshadRealtimeAdaptiveEstimation2024} and qubit control within a noisy environment \cite{cappellaroSpinbathNarrowingAdaptive2012, shulmanSuppressingQubitDephasing2014, scerriExtendingQubitCoherence2020, berrittaRealtimeTwoaxisControl2024, berrittaEfficientQubitCalibration2025}. In all applications involving experiment adaptation and feedback, it is crucial to minimize the processing time required to update the probability distribution according to Bayes rule, {and generate the optimal control sequence for the next measurements}. In quantum sensing, the processing time  {is ideally} negligible compared to the data acquisition time, or at least the shortening of sensing time given by real-time adaptation needs to be larger than the  computational time invested to achieve this. 

Despite its success for small-scale quantum technologies, the extension of Bayesian approaches to learn highly multi-dimensional quantum systems is highly non-trivial, since exact Bayesian inference is often computationally intractable. As the parameter space grows, evaluating and sampling from the posterior becomes exponentially more expensive, in what is known as the ``curse of dimensionality'' \cite{bellman_DynamicProgramming_1984, bengtsson_CurseofdimensionalityRevisitedCollapse_2008}. This has restricted the application of Bayesian inference to experiments in quantum tomography \cite{huszarAdaptiveBayesianQuantum2012, kravtsovExperimentalAdaptiveBayesian2013, lu_BayesianTomographyHighdimensional_2022, chapman_BayesianHomodyneHeterodyne_2022} and quantum sensing \cite{bonatoOptimizedQuantumSensing2016, santagatiMagneticFieldLearningUsing2019, valeriExperimentalAdaptiveBayesian2020, joasOnlineAdaptiveQuantum2021, daurelioExperimentalInvestigationBayesian2022, valeriExperimentalMultiparameterQuantum2023, arshadRealtimeAdaptiveEstimation2024, belliardoOptimizingQuantumenhancedBayesian2024} involving only a handful of parameters \footnote{We stress that this exponential {growth} is not related to the exponential complexity in time and memory of simulating and certifying quantum systems, which we are not trying to address in this paper.}. 

Here we introduce a powerful and practical  solution to this challenge, deploying an approximate approach that extends the applicability of Bayesian inference to high-dimensional learning problems in quantum settings. To avoid the intractable scaling of exact Bayesian techniques for  large numbers of parameters, we deploy variational Bayesian inference (VBI) \cite{blei_VariationalInferenceReview_2017}, introducing a family of simpler distributions and selecting the one that best approximates the true posterior. In Section \ref{sec:VBI_learning}, we show that, as the dimensionality of the estimation problem grows, we can choose distributions whose sampling time and memory complexity scale polynomially with the number of parameters, enabling us to efficiently estimate several tens of parameters.

In addition to enabling the estimation of large numbers of parameters, our approach is also applicable to cases where the system model is not uniquely defined, for example for quantum systems coupling to an environment with {a} number of constituents unknown {a priori} (e.g.~a spin bath). 
In Section \ref{sec:model_learning}, we show that, for cases where the possible model classes are organized in a hierarchical structure with models higher in the hierarchy featuring higher complexity, our framework successfully identifies the model with minimal complexity that explains the data. This is for example the case of Hamiltonian or Lindbladian identification where the most complex models contain a large number of terms with their respective couplings and parameters, suggesting the need {for} a form of ``Occam{'s} razor'' heuristic \cite{gentileLearningModelsQuantum2021, wallace_LearningDynamicsMarkovian_2025, fioroniLearningAgentbasedApproach2025} privileging simple models {that} fit the data. A Bayesian approach provides explainable insights into the physics underpinning the observations, highlighting the constituent sub-systems, the involved physical processes, and the corresponding coupling rates: this is generally less clear with deep learning strategies to learn models for the generators of dynamics \cite{mazza_MachineLearningTimelocal_2021, cemin_InferringInterpretableDynamical_2024, schorling_MetalearningCharacteristicsDynamics_2025}.

The general framework we introduce here can be applied to a variety of quantum systems. In Section \ref{sec:nanonmr_example}, we consider the problem of identifying individual nuclear spins by a quantum sensor consisting of a single electronic spin. This example was chosen for its complexity, with the identification of several tens of individual {spins} demonstrated in experiments \cite{vandestolpeMapping50spinqubitNetwork2024}, and its practical relevance in both developing nanoscale magnetic resonance techniques \cite{budakianRoadmapNanoscaleMagnetic2024} and quantum networks of nuclear spin qubits \cite{bradleyTenQubitSolidStateSpin2019, randallManybodylocalizedDiscreteTime2021}. More specifically, this application shows the power of our proposed framework by addressing a complex problem with a nonlinear signal, a number of spins not known in advance (as it depends on the random environment that each individual spin defect experiences), and a highly multidimensional parameter space. We demonstrate in simulation how our framework can perform model selection using a variational Bayesian inference approach, reliably identifying $15$ $^{13}$C spins (corresponding to $30$ hyperfine parameters) with only a fraction of the total data collected in the laboratory. In processing the experimental data we use an ansatz for the Bayesian posterior distribution able to describe up to $40$ individual spins, and we show how we are able to compete with state-of-the-art deep learning approaches~\cite{jungDeepLearningEnhanced2021, varona-uriarteAutomaticDetectionNuclear2024} with limited experimental data and training time. Our Bayesian approach generates an approximate posterior on time scales of a few minutes, and does not assume prior knowledge of the settings used to acquire the input signal: this makes it  perfectly compatible with the future integration of real-time sequential Bayesian design, with the goal to minimize the characterization/sensing time.

\begin{figure*}[!htbp]
    \centering
\includegraphics[width=1.0\textwidth]{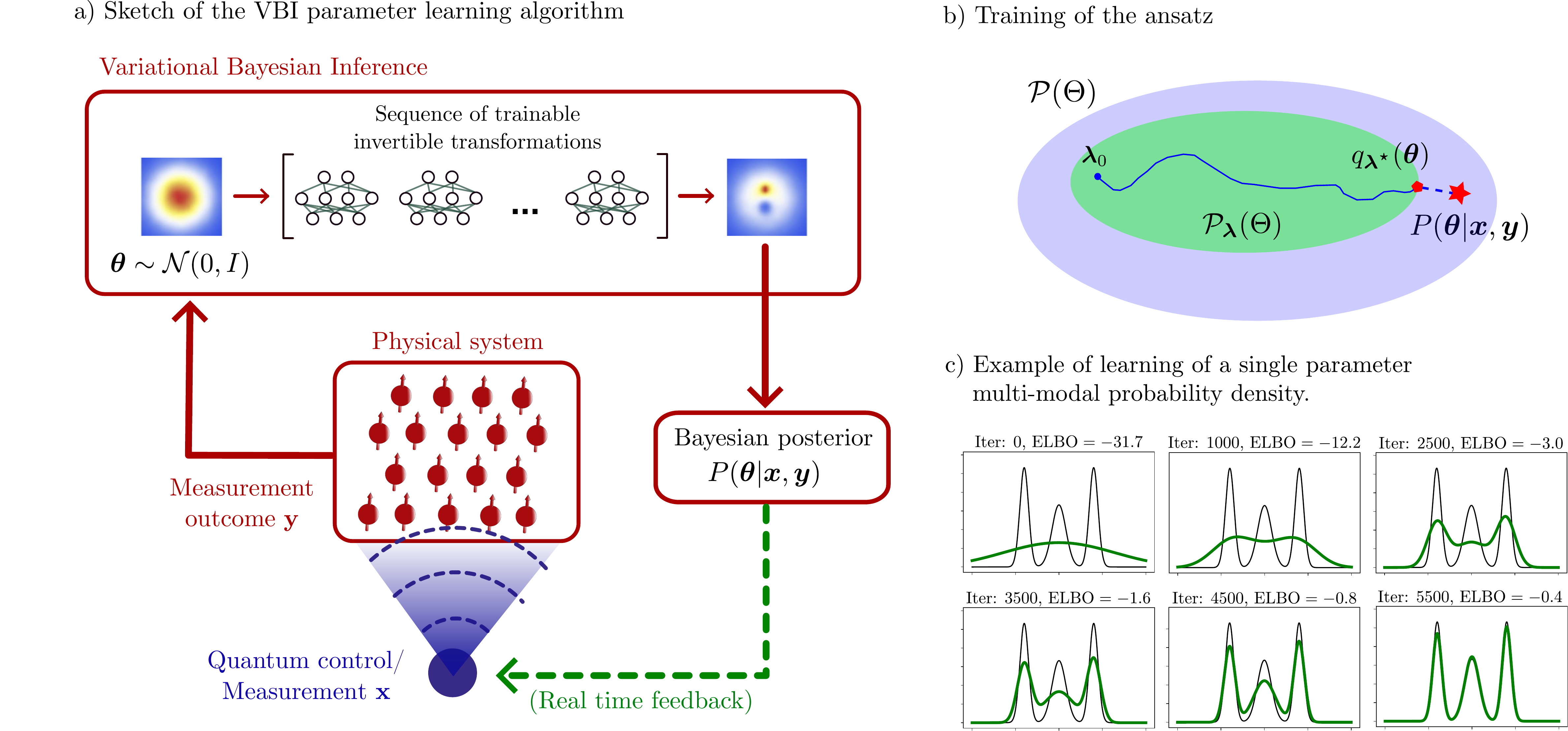}
    \caption{\justifying \textbf{Introduction to high-dimensional parameter estimation for quantum experiments using variational Bayesian inference (VBI).} \textbf{a)} Schematic representation of a quantum estimation experiment based on VBI. The physical system, here represented by an ensemble of spins, undergoes a series of measurements, controlled by the parameters $\vx$, obtaining the outcomes $\vy$. This information is used to train the posterior ansatz in VBI, i.e. a function that transforms $\vtheta \sim \mathcal{N} (0, \mathds{I})$ in samples extracted from the posterior distribution for $\vtheta$, that contains the information on the batch of measurement just executed. The Bayesian posterior can be used in a real-time feedback algorithm to calculate the controls for the next batch of measurements, adaptively. \textbf{b)} The space of distributions spanned by varying $\vlambda$ is indicated by the green set, and it corresponds to $\mathcal{P}_{\vlambda} (\Theta) := \lbrace q_{\vlambda} (\vtheta)  \mid  \vlambda \in \mathbb{R}^k \rbrace$ which is a subset of the space of all possible probability densities over $\Theta$, indicated with $\mathcal{P} (\Theta)$. The training of the ansatz is a procedure that starting from an initial distribution identified by $\vlambda_0$, through a sequence of update steps of the values of $\vlambda$ finds $\vlambda^\star$ which for which $q_{\vlambda^\star} (\vtheta)$ minimizes the Kullback-Leibler divergence with the true posterior. The point $\vlambda^\star$ is indicated by the red dot, while the posterior is the red star in the picture. \textbf{c)} Example of application of VBI to a multi-modal single-parameter distribution. The algorithm uses a mixture of Gaussians as ansatz. Maximization of the ELBO (Eq. \ref{eq:elbo_definition}), over an increasing number of steps, minimizes the Kullback-Leibler divergence with the true distribution until a sufficiently good approximation is reached, using the means and standard deviations of the three Gaussian as training parameters. More complex distributions can be approximated by a sequence of invertible transformation, for example implemented by neural networks \cite{huangNeuralAutoregressiveFlows2018}(e.g. normalized flow). In the plot, the ground truth distribution is shown in black, and the green lines correspond to the VBI approximations. The number on the top left of each sub-plot corresponds to the iterations index during training.}
    \label{fig:intro}
\end{figure*}

\section{Variational Bayesian inference for quantum estimation}
\label{sec:VBI_learning}
In this section, we first discuss how to construct the outcome probability distributions for parameter estimation in quantum experiments. We will then introduce variational Bayesian inference, and compare it to other approaches used in quantum metrology, such as particle filters and Markov Chain Monte Carlo (MCMC), in terms of scalability and precision.

\subsection{Bayesian formulation of the estimation problem}
\label{subsec:channel_state_est}
The goal of quantum parameter estimation is to determine the value of $\vtheta \in \Theta$ characterizing a completely positive, trace-preserving (CPTP) linear map~\cite{nielsenQuantumComputationQuantum2012, holevoQuantumSystemsChannels2019}, which we denote $\mathcal{E}_{\vtheta} (\cdot)$. This map corresponds to a physical quantum channel, which describes the (possibly non-unitary) evolution of an input quantum system. In order to complete the estimation task, a probe state $\rho (x)$ is evolved through the quantum channel, producing an encoded state $\rho (x, \vtheta) := \mathcal{E}_{\vtheta} (\rho (x) )$. In this manuscript, consistently with the notation adopted in~\cite{belliardoModelawareReinforcementLearning2024, belliardoApplicationsModelawareReinforcement2024}, we use $x \in \mathcal{X}$ to indicate the controls available to the experimenter. The goal is to estimate $\vtheta$ from $M$ measurements performed on the set of states $\lbrace \rho (x_t, \vtheta) \rbrace_{t=1}^M$, and we assume that each encoded state can feature a different set of controls ($x_t$) for each experiment. 

In the case of state estimation, we are given $\rho (\vtheta)$ directly, without the possibility of manipulating the probe state. After this encoding stage, a positive operator-valued measurement (POVM) is performed on the encoded state. Motivated by experimental feasibility, we assume that only a single copy of the encoded state is available for each measurement. 

A generalized measurement on $\rho(x, \vtheta)$ is represented by the set of operators $\lbrace {E}{(x, y)} \rbrace_{y}$ fulfilling the normalization condition $\sum_{y} {E}(x, y)=\mathds{1}$. The symbol $y$ indicates the outcome of {the} measurement, while  $x$ represents  {collectively all the} control parameters available to the experimenter at the measurement stage. In the following, we will use a single symbol $x_t$ to indicate both the controls on the state preparation and the measurement in an experiment identified by the index $t \in \lbrace 1, \cdots, M \rbrace$.  Throughout this manuscript, we use the term ``experiment'' to indicate a single step of encoding and measurement, while the set of all experiments carried out to estimate the same $\vtheta$ on the quantum channel, together with the data processing, is called the ``estimation''. Given the whole set of $M$ experiments, we indicate with $\vx_M := (x_1, x_2, \cdots, x_{M})$ the complete list of controls and with $\vy_M := (y_1, y_2, \cdots, y_{M})$ the list of corresponding outcomes. 

According to Born's rule, the probability outcome for $y_t$ is given by
\begin{equation}
	p(y_t \mid  x_t, \vtheta) := \mbox{Tr}[ {E} (x_t, y_t) {\rho} (x_t, \vtheta) ] \; .
	\label{eq:bornrule}
\end{equation}
We label $p(y_t \mid x_t, \vtheta)$ as \textit{the model}.  The experimental data {are} processed through Bayes' formula to get the posterior probability of $\vtheta$, i.e.
\begin{equation}
	P( \vtheta \mid \vx_M, \vy_{M}) := \frac{\pi (\vtheta) \prod_{t=1}^{M} p(y_t \mid x_t, \vtheta)}{\int d \vtheta \pi (\vtheta) \prod_{t=1}^{M} p(y_t \mid x_t, \vtheta)}.
	\label{eq:bayesian}
\end{equation}
The Bayesian approach, in addition to enabling uncertainty quantification from the posterior, is efficient in the use of data,  and allows seamless incorporation of prior {beliefs} about the model and its parameters, which are indicated in the above formula with the distribution $\pi(\vtheta)$. 

We will further use the following definitions and nomenclature. The joint likelihood is defined as
\begin{equation}
    p(\vy_M \mid \vx_M, \vtheta) := \prod_{t=1}^{M} p(y_t \mid x_t, \vtheta) \; ,
\end{equation}
and
\begin{equation}
    P(\vy_M \mid \vx_M) := \int d \vtheta \pi (\vtheta) \prod_{t=1}^{M} p(y_t \mid x_t, \vtheta) \; ,
\end{equation}
is called the ``evidence'' or marginal likelihood.

\subsection{Computational approaches}
\label{subsec:computational_bayes}

Unless both prior and model fit within specific classes of distributions for which Bayes' rule can be treated analytically, e.g. the exponential family and conjugate priors, the integral in the denominator of Eq.~\eqref{eq:bayesian} can only be computed numerically. We focus on four options to apply {Bayes'} rule numerically.

The first one, based on \textit{sequential Monte Carlo} (SMC) and particle filtering \cite{gordon_NovelApproachNonlinear_1993, doucet_SequentialMonteCarlo_2001, delmoral_SequentialMonteCarlo_2006}, is very popular for quantum estimation experiments {that are} executed with few parameters \cite{granadeRobustOnlineHamiltonian2012, wiebeHamiltonianLearningCertification2014, valeriExperimentalAdaptiveBayesian2020,joasOnlineAdaptiveQuantum2021, gentileLearningModelsQuantum2021, valeriExperimentalMultiparameterQuantum2023, belliardoApplicationsModelawareReinforcement2024, belliardoOptimizingQuantumenhancedBayesian2024, belliardoModelawareReinforcementLearning2024}. These computational techniques, based on discretizing the parameter space, do not scale well to high-dimensional systems, as the memory and operations required for the Bayesian update scale exponentially in the number of parameters. Re-sampling techniques \cite{gordon_NovelApproachNonlinear_1993, kitagawa_MonteCarloFilter_1996, liu_SequentialMonteCarlo_1998, liu_CombinedParameterState_2001, cappe_OverviewExistingMethods_2007,  li_ResamplingMethodsParticle_2015, murray_ParallelResamplingParticle_2016} are useful in mitigating this exponential dependence, but the computational costs remain prohibitive for  {dozens of} parameters. {SMC can readily incorporate real-time experiment design, as the posterior is updated online after each new measurement almost instantly.}

A second popular computational approach is the use of \textit{Markov Chain Monte Carlo} \cite{gelfand_SamplingBasedApproachesCalculating_1990, brooks_HandbookMarkovChain_2011, jones_MarkovChainMonte_2022}(MCMC)  to approximate the posterior as the steady-state distribution of a Markov process~\cite{wallace_LearningDynamicsMarkovian_2025, poteshman_HighthroughputSpinbathCharacterization_2025}. MCMC techniques present a favorable cost scaling in the number of studied parameters, and allow the extension of quantum estimation to high-dimensional problems. Their main limitation is that the Markov chain only  faithfully represents the posterior distribution when it has properly thermalized, which  {can} take a significant time (burn-in).
The  {burn-in} time cannot be shortened by training multiple chains in parallel, so that even today's GPUs have limited impact on improving the speed of this technique, which is sequential in nature. 

{A third popular option is \textit{nested sampling} (NS) \cite{skilling_NestedSampling_2004, skilling_NestedSamplingGeneral_2006, ashton_NestedSamplingPhysical_2022} which transforms the multi-dimensional posterior integration problem into a one-dimensional evidence calculation by iteratively sampling from prior-constrained shrinking likelihood contours. This method excels at efficiently exploring multi-modal posteriors and computing Bayesian model evidence, and can offer a more favorable scaling for high-dimensional problems compared to plain MCMC. In a similar fashion to MCMC, nested sampling implementations might struggle with real-time updates due to sequential sampling needs.}

The fourth computational approach is \textit{approximate Bayesian inference}, in particular variational Bayesian inference (VBI) \cite{peterson_MeanFieldTheory_1987, hinton_KeepingNeuralNetworks_1993, neal_ViewEmAlgorithm_1998, saul_MeanFieldTheory_1996, jaakkola_VariationalApproachBayesian_1997, blei_VariationalInferenceReview_2017}. This algorithm prescribes the use of a trainable \textit{ansatz}, often based on tailored neural networks (NN), to represent the posterior distribution. Although the update of the NN weights in the training process remains sequential, using large mini-batches  {of sampled parameters and training data} at each step can be a very effective way of exploiting the parallelization offered by current GPU hardware. The fast estimation of an approximation {to} the actual posterior distribution makes real-time adaptive experiments possible.

These reasons motivate the choice {to} deploy VBI as a computational approach to target parameter estimation in highly multidimensional quantum experiments, {with many data points to process}. The next subsection briefly recalls the working principles of VBI. {See also Fig.~\ref{fig:comparison_filtering} for a comparison of the four techniques on a toy model involving multi-frequency estimation.}

\subsection {Variational Bayesian inference}
\label{sec:bayesian_learning}
VBI assumes an ansatz probability density $q_{\vlambda}(\vtheta)$, where $\vlambda$ are trainable parameters, and it aims to minimize the Kullback-Leibler divergence (measuring discrepancy between distributions) between the ansatz and the true posterior, i.e. $\text{KL} (q_{\vlambda} (\vtheta) \parallel  P(\vtheta \mid  \vx_M, \vy_M))$. However, it is not possible to directly minimize this divergence, since it depends on the posterior whose normalization is typically unknown. It is instead possible to split the dependence of this divergence between $\vlambda$, i.e. the training variables, and the unknown normalization of the posterior, i.e. the evidence, as in the following~\cite{kingma_AutoEncodingVariationalBayes_2022, blei_VariationalInferenceReview_2017},
\begin{widetext}
    \begin{equation}
    \begin{aligned}
    \text{KL} (q_{\boldsymbol{\lambda}} (\vtheta) & \parallel P(\vtheta \mid  \vx_M, \vy_M)) 
    := - \mathbb{E}_{\vtheta \sim q_{\boldsymbol{\lambda}}} \left[ 
    \log \frac{P(\vtheta \mid  \vx_M, \vy_M)}{q_{\boldsymbol{\lambda}} (\vtheta)} 
    \right] \\
    &= - \mathbb{E}_{\vtheta \sim q_{\boldsymbol{\lambda}}} \Bigg[ 
    \log \pi (\vtheta) 
    + \sum_{t=1}^{M} \log p(y_t \mid x_t, \vtheta) 
    - \log P(\vy_M \mid \vx_M) 
    - \log q_{\boldsymbol{\lambda}} (\vtheta) 
    \Bigg] \\
    &= - \mathbb{E}_{\vtheta \sim q_{\boldsymbol{\lambda}}} \Bigg[ 
    \log \pi (\vtheta) 
    + \sum_{t=1}^{M} \log p(y_t \mid x_t, \vtheta) 
    - \log q_{\boldsymbol{\lambda}} (\vtheta) 
    \Bigg] + \log P(\vy_M \mid \vx_M) \\
    &:= - \text{ELBO}(\boldsymbol{\lambda}) + \log P(\vy_M \mid \vx_M) \; ,
    \end{aligned}
    \label{eq:elbo_definition}
    \end{equation}
\end{widetext}
where we have introduced the evidence lower-bound (ELBO). This quantity takes its name from the property $\text{ELBO}(\boldsymbol{\lambda}) \le \log P(\vy_M  \mid  \vx_M)$, which follows from $\text{KL} (q_{\boldsymbol{\lambda}} (\vtheta) \parallel P (\vtheta \mid  \vx_M, \vy_M)) \ge 0$. Instead of directly minimizing the KL divergence, VBI seeks to maximize the ELBO.

The natural candidate for the loss function to be minimized in the training is $- \text{ELBO}(\vlambda)$, since it has the same gradient of the KL-divergence:
\begin{equation}
    \frac{\partial}{\partial \boldsymbol{\lambda}} \text{KL}\big(q_{\boldsymbol{\lambda}}(\boldsymbol{\theta}) \parallel P(\boldsymbol{\theta} \mid \boldsymbol{x}_t, \boldsymbol{y}_t)\big) = - \frac{\partial \text{ELBO}(\boldsymbol{\lambda})}{\partial \boldsymbol{\lambda}} \; .
\end{equation}
The next question to address is how to construct the ansatz. Assuming $\vtheta$ consists of continuous parameters, the ansatz $q_{\vlambda} ( \vtheta)$ can be constructed starting from a set of normally distributed variables $\vz \sim \mathcal{N} \left(0, \mathds{I}_{d}\right)$, which are transformed through $\vtheta := f_{\vlambda} (\boldsymbol{z})$, called a normalizing flow~\cite{rezendeVariationalInferenceNormalizing2016} ($d$ is the dimensionality of $\vtheta$). If $f_{\vlambda}$ is invertible, the transformed probability distribution for $\vtheta$ is given by 
\begin{equation}
    q_{\boldsymbol{\lambda}}(\vtheta) \, := \, \mathcal{G}(\boldsymbol{z}(\vtheta)) \Bigl\vert  \det \frac{\partial f_{\vlambda}}{\partial \boldsymbol{z}}(\boldsymbol{z}(\vtheta)) \Bigr\vert ^{-1},
    \label{eq:first_normalizing_flow}
\end{equation}
where we have indicated with $\mathcal{G} (\boldsymbol{z})$ the probability density of a multivariate Gaussian and $\boldsymbol{z}(\vtheta) := f_{\vlambda}^{-1}(\vtheta)$. Crucially, we need both operations, i.e. sampling from the ansatz and computing the density $q_{\vlambda} (\vtheta)$, to be easy to perform in order to compute the ELBO. If we choose $f_{\vlambda}$ to be the affine transformation $\vtheta = f_{L, \boldsymbol{\mu}}(\boldsymbol{z}) := L \boldsymbol{z} + \boldsymbol{\mu}$, with trainable $L$ (lower triangular matrix and invertible) and $\boldsymbol{\mu}$, the ansatz will span the space of all multivariate Gaussian probability densities, when $\vmu$ and $L$ are varied. To further extend the class of parametrized distributions, and account for non-gaussianities, we include a series of non-linear transformation before $f_{L, \boldsymbol{\mu}}$, so that the complete transformation $f(\boldsymbol{z})$ is given by
\begin{equation}
    f(\boldsymbol{z}) := f_1 \circ f_2 \circ \cdots f_K \circ f_{L, \boldsymbol{\mu}}  (\boldsymbol{z})\; .
\end{equation}
These nonlinear transformations can be implemented as neural autoregressive layers ~\cite{huangNeuralAutoregressiveFlows2018}, and are defined to be invertible and such that the corresponding log-determinant needed in the ELBO can be computed efficiently. Note that this choice is not exclusive and it is also possible to choose other neural architectures, or simply choose $q_{\boldsymbol{\lambda}}(\vtheta)$ as a classical family of parametric (exponential) distributions. For instance, in the next subsection, $q_{\boldsymbol{\lambda}}(\vtheta)$ is chosen to be a multivariate Gaussian density with diagonal covariance matrix.

\subsection{Scalability of the Bayesian estimation procedure}
\label{subsec:scalability}
\begin{figure*}[!htbp]
    \centering
\includegraphics[width=1.0\textwidth]{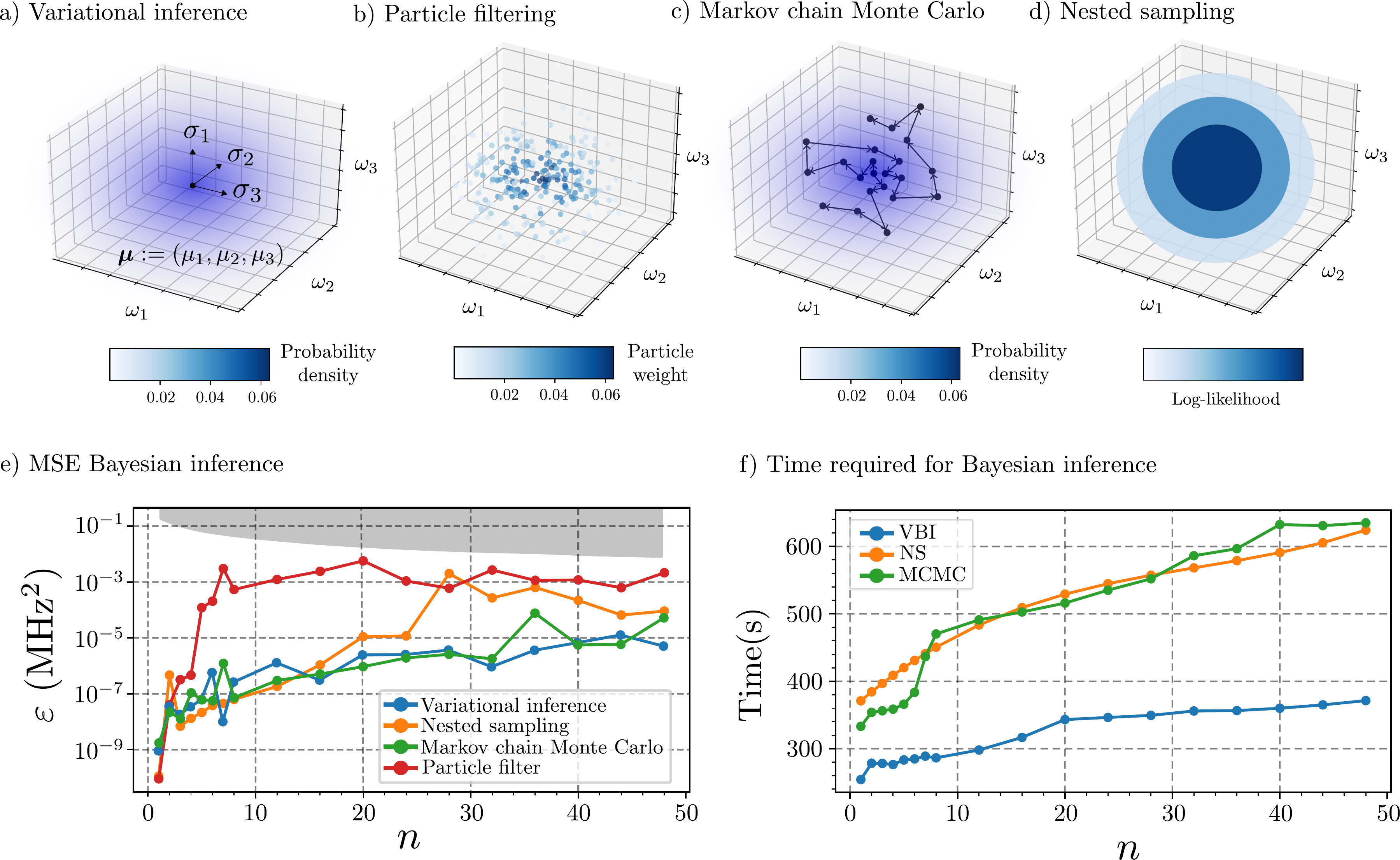}
    \caption{\justifying \textbf{Scalability of VBI for multi-dimensional parameter estimation.} Schematic representation of a $3$D Bayesian posterior distribution for  three frequencies ($\omega_1, \omega_2, \omega_3$), estimated with four different techniques: \textbf{a)} VBI based on a Gaussian ansatz, which depends on $6$ parameters only: the three components of the position of the Gaussian $\mu_1, \mu_2, \mu_3$ and the standard deviations $\sigma_1, \sigma_2, \sigma_3$; \textbf{b)} particle filtering, with an ensemble of particles used to approximate the posterior (in the figure, the color of the particles represents their weight); \textbf{c)} Markov Chain Monte Carlo (MCMC),  with the posterior distribution represented by an ensemble of samples obtained sequentially from a thermalized Markov chain, which has the Bayesian posterior as fixed point distribution; \textbf{d)} nested sampling, which evaluates the volume of the region of parameters with a given log-likelihood, and reduces the computation of the multidimensional evidence in Bayesian inference to a 1D integral. \textbf{e)} Comparison of the mean squared error $\varepsilon$ in the estimation of $n$ frequencies (Eq.~\eqref{eq:square_error}) with different Bayesian estimation algorithms(particle filter: red line;  Markov Chain Monte Carlo: green line; nested sampling: orange line; VBI: blue line), all operating on the same dataset. The error refers to a single instance of the estimation task, we do not report the statistical fluctuations on the plot. The shaded region corresponds to the error expectation value achievable without any measurements being performed, but guessing at random the frequencies according to the prior: this represent a limit in the performance of all estimation algorithms. \textbf{f)} Comparison of run time for VBI, nested sampling and MCMC, on the same problem discussed in (e). The results show that VBI scales better with increasing problem dimensionality than competing approaches. All the simulations have been performed on the CPU. Particle filtering is not considered here as it is generally fast but, as evidenced in (e), it is not competitive in terms of estimation error in the high-dimensional regime. The results in (e) and (f) support the claim that VBI can implement Bayesian inference on a high-dimensional problem with the possibility to include real-time adaptivity}
\label{fig:comparison_filtering}
\end{figure*}

In this subsection, we discuss an example of how variational Bayesian inference outperforms particle filtering as the number of parameters in the model increases. We consider a simple toy model where multiple frequencies are encoded on a probe quantum state when interacting with an environment, through the CPTP map:
\begin{equation}
    \mathcal{E}_{\vomega} (\cdot) = \frac{1}{n} \sum_{i=1}^n U_{\omega_i} \cdot U_{\omega_i}^\dagger \; 
    \label{eq:encoding}
\end{equation}
The goal is to estimate the vector of unknown frequencies $\vomega := (\omega_1, \omega_2, \cdots , \omega_n)$, see Fig.~\ref{fig:comparison_filtering}. This corresponds to a Ramsey experiment with a single qubit, initially prepared in the state $\ket{\psi} = \left( \ket{0}+\ket{1}\right)/\sqrt{2}$, which evolves freely in the environment for a time $\tau$, according to unitary operators $U_{\omega_i} := e^{-\frac{i}{2} \tau \omega_i \sigma_z}$. A qubit measurement in the superposition basis outputs the outcomes $\ket{\pm1}$, with probabilities, respectively, given by
\begin{equation}
    p(\pm 1 | \tau, \vomega) := \frac{1}{2} \pm \frac{1}{2 n} \sum_{i=1}^n \cos (\omega_i \tau) \; ,
    \label{eq:simplified_phase_model}
\end{equation}
The frequencies are randomly selected in the range $\omega_i \in [0, 1] \MHz$ uniformly. We solve this multidimensional estimation problem for an increasing number of frequencies, {comparing VBI with different Bayesian estimation techniques (MCMC, nested sampling, and particle filtering).} The particle filter evaluates the probability distributions over a number $N_p=16384$ of points (known as ``particles'') in the multi-dimensional parameter space, implemented as detailed in~\cite{belliardoModelawareReinforcementLearning2024, belliardoApplicationsModelawareReinforcement2024}.
{
We have chosen the dataset and the number of parameters $n$ such that the size of the problem is similar to the estimation of hyperfine parameters discussed in Sec.~\ref{sec:nanonmr_example}. The dataset is composed of $M=4096$ points with different $\tau$, selected in such way that $\log \left( \tau / 1 \mus \right) $ is sampled uniformly in $[-1, +5]$. This choice is motivated by the error analysis of phase estimation in~\cite{belliardo_AchievingHeisenbergScaling_2020}. For each value of $\tau$, $R=1024$ experiments with a binary outcome are simulated.} 

Nested sampling is based on evaluating the volumes of the shells of parameter space with a given log-likelihood. For this comparison we have implemented adaptive shrinking of the prior volume and replacement of low-likelihood points through MCMC and slice sampling under likelihood constraints~\cite{neal_SliceSampling_2003}. The algorithm features adaptive proposal distributions, likelihood-constrained slice sampling, and dynamic live point populations, which are all standard improvements over the baseline algorithm~\cite{higson_DynamicNestedSampling_2019}. The initial and maximum number of live points are respectively $20$ and $500$, we have simulated $4096$ iterations, with $500$ MCMC steps used, inside the NS algorithm, to sample new live points in each iteration. The last error in Fig.~\ref{fig:comparison_filtering} refers to inference using MCMC, which is based on sampling from a Markov chain whose equilibrium distribution is the Bayesian posterior. We have implemented adaptive Metropolis-Hastings MCMC, which generates posterior samples via random walk proposals from a multivariate normal~\cite{haario_AdaptiveMetropolisAlgorithm_2001}. We simulate $3 \cdot 10^6$ steps of a single Markov chain, with $10\%$ burn-in. As described above, VBI does not evaluate the posterior over a set of particles {or through sampling}, but approximates it with a set of simpler functions chosen as ansatz. Here we use the ``mean field'' ansatz, i.e. we assume the parameters are uncorrelated and we approximate the posterior with a product of Gaussians. The trainable parameters of the posterior are therefore the mean values $\mu_i$ and the standard deviations of each Gaussian $\sigma_i$, i.e. for $n$ frequencies the number of trainable parameters scales polynomially as $2n$.
Labeling the Bayesian estimator for $\omega_i$ as $\hat{\omega}_i$, we compute the error in a single estimation as
\begin{equation}
    \varepsilon := \frac{1}{n} \sum_{i=1}^n (\hat{\omega}_i - \omega_i)^2 \; ,
    \label{eq:square_error}
\end{equation}
where we choose to order the list of frequencies and their estimators in ascending order, i.e. $\omega_1 < \omega_2 < \cdots < \omega_n$ and $\hat{\omega}_1 < \hat{\omega}_2 < \cdots < \hat{\omega}_n$. 

Fig.~\eqref{fig:comparison_filtering} compares the error $\varepsilon$ as a function of the number of frequencies {for all the techniques described here}. {The shaded gray region is the average error per frequency achievable when no measurements are carried out. This is not a constant; on the contrary, it slightly decreases as the number of estimated frequencies grows}. This effect is due to the permutation invariance of the frequency labels and can be understood as follows.  For an arbitrary frequency $\omega$, if no measurements are performed, each of the $\widehat{\omega}_i$ estimators is selected randomly and uniformly in the parameter domain. The larger the number of unknown frequencies $n$, the higher the probability that one of the $n$ estimators $\hat{\omega}_i$ will randomly fall close to the selected $\omega$. Therefore, we expect the average error to decrease with the number of frequencies.
{We observe that VBI with mean field ansatz is as good as MCMC, but it has a more favorable time scaling.} %

\begin{figure*}[!htbp]
    \centering
\includegraphics[width=1\textwidth]{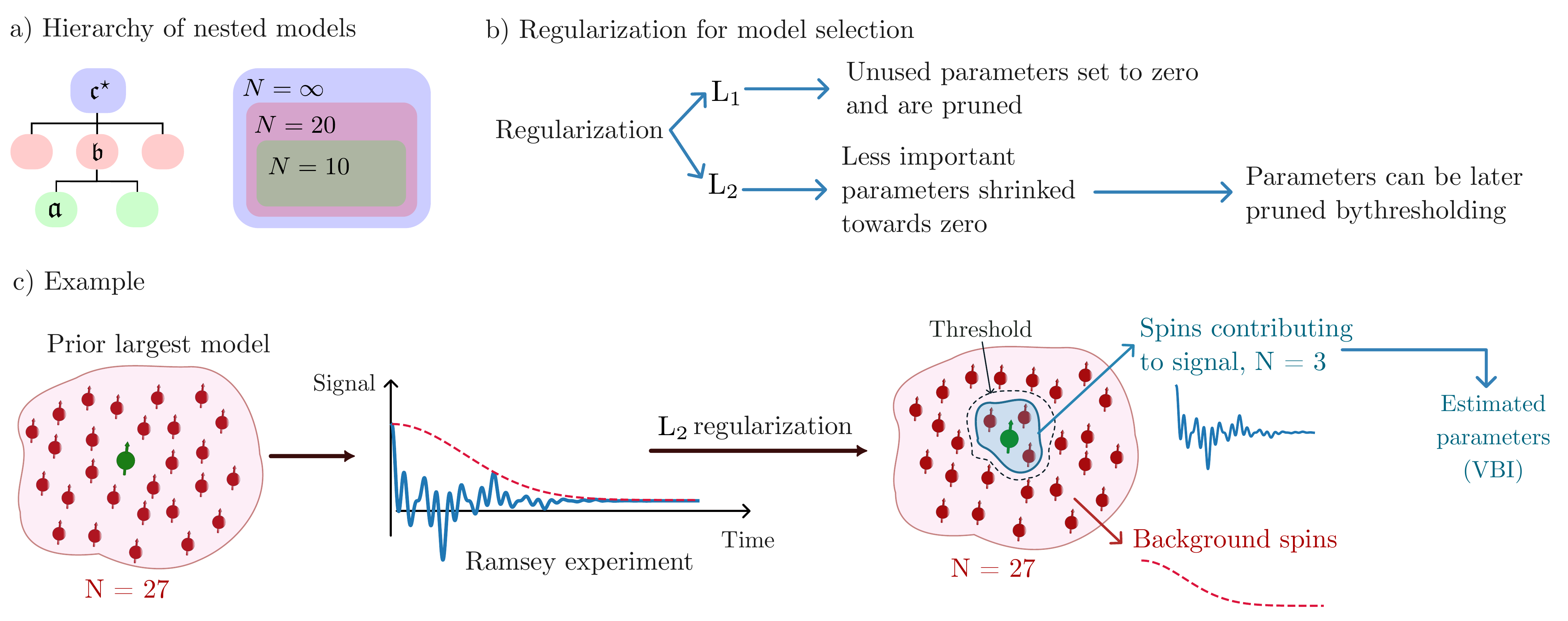}
    \caption{\justifying \textbf{Model selection.} \textbf{a)} On the left, we show a schematic representation of a hierarchical set of model, where $\class{a}$ is a sub-model of $\class{b}$, which is a sub-model of $\class{c}^\star$. On the right, we represent the nested models of an environment constituted of different numbers $N$ of qubits. \textbf{b)} The regularization procedure described in Section \ref{subsec:minimal_model_selection} enables selecting the simplest model that explains the data. L$_1$ regularization (Eq. \ref{eq:L1_laplace}) enforces sparsity, setting the parameters that are not important exactly to zero, thereby pruning them from the model. L$_2$ regularization (Eq. \ref{eq:L2_gaussian}) results in values close to zero (but not exactly zero) for parameters which impact the outcome less: these can be then pruned in post-processing by imposing a threshold. \textbf{c)} Here we show an example of L$_2$ regularization. We consider a set of models with a central spin and, at maximum, $N=27$ spins in the surrounding bath. We then perform a Ramsey experiment on the central spin, in which the central spin is prepared in an equal superposition and let freely evolve under the influence of the bath. The outcome signal consists of multiple decaying oscillations, corresponding to the distribution of magnetic field values given by the different spin configurations. VBI with L$_2$ regularization learns the coupling parameters to the 27 environmental spins: in this case, 24 of them are close to zero. A threshold separates the $3$ spins that contribute to the oscillating signal from the $24$ background spins that contribute to the decaying envelope (corresponding to the $T_2^*$ dephasing time). If we are both estimating $T_2^*$ and all the spins in the environment, we could chose to use the L$_1$ regularization, set the coupling of the background spins exactly to zero in the ansatz and explain the envelop with a non-zero $1/T_2^*$. In Section \ref{sec:nanonmr_example}, we consider a similar setting with a different experiment, using both $(T_2^\star)$ and the background spins.}
    \label{fig:model_selection}
\end{figure*}

\section{Learning models for quantum systems using variational Bayesian inference}
\label{sec:model_learning}
In this section we discuss the selection of a model within a set of possible models using VBI, as summarized in Fig.~\ref{fig:model_selection}.

\subsection{Model selection for a set of nested models}
\label{subsec:model_selection}
In the previous section, we have discussed the application of Bayesian inference to the estimation of a fixed number of parameters. We will {now} extend the discussion to model selection, to describe cases where there might be a number of different descriptions of a given quantum system, each with a different number of parameters, but all compatible with the observed data. This is useful, for example, to characterize the interaction of a quantum system with its environment, whose composition might not be precisely known. Examples of this situation are a central electron spin interacting with a number of nuclear spins, where both the number and couplings of the spins are unknown {a priori} (discussed in {detail} in Section \ref{sec:nanonmr_example}), or a superconducting qubit coupled to a bath of two-level fluctuators \cite{burnett_EvidenceInteractingTwolevel_2014, degraaf_TwolevelSystemsSuperconducting_2020, lucas_QuantumBathSuppression_2023, liu_ObservationDiscreteCharge_2024}.

We assume the quantum system can be described by a discrete set of classes $\mathcal{C}$ (possibly infinite). Each class $\class{c} \in \mathcal{C}$ is described by a set of continuous parameters $\vtheta_\class{c} \in \Theta_\class{c}$, possibly with a different number of parameters for each class $\class{c}$. 

The model selection problem involves finding the class $\class{c}$ and the values of the associated parameters $\vtheta_\class{c}$ that explain the experimental data. To achieve this, instead of a single posterior distribution $P(\vtheta \mid \vx_t, \vy_t)$, we need to specify the list of marginal posteriors conditioned on the model's class, i.e. $\lbrace P(\vtheta_\class{c}  \mid \class{c}, \vx_t, \vy_t) \rbrace_{\class{c} \in \mathcal{C}}$, and {the} list of class probabilities $\lbrace P(\class{c} \mid \vx_t, \vy_t) \rbrace_{\class{c} \in \mathcal{C}}$. This increased complexity affects the application of {Bayes'} rule, which in this case needs to account for both the posterior in the continuous parameters of each model and the probability over $\mathcal{C}$.

We assume that among different classes of models there is a natural partial ordering, defined by identifying a model as {a} sub-model of a more general class (Fig.~\ref{fig:model_selection}(a)). For example, a system with $M > N$ qubits is more general than the model with $N$ qubits (and includes it as a particular case), when the couplings of the remaining qubits are set to zero. 

Within this formalism, we assume we can construct a pruning function which, given a class $\class{b} \in \mathcal{C}$ and the continuous parameters $\vtheta_{\class{b}}$, returns the class $\class{a} \in \mathcal{C}$ and its associated parameters such that $\class{a}$ is a sub-model of $\class{b}$ that has been pruned of all the zero parameters $(\vtheta_{\class{b}})_i = 0$. For example, given a model with $M$ qubits, if $M - N$ qubits have zero couplings, then the pruning of these non-interacting qubits gives an equivalent model with $N$ qubits. 

In other words, we assume the existence of a class $\class{c}^\star \in \mathcal{C}$ such that we can view all models as its subclasses, after setting some of the parameters to zero, i.e. we can reduce the problem of model learning to parameter estimation for $\vtheta_{\class{c}^\star}$, where the goal is now both to minimize the number of non-zero elements of the vector of continuous parameters and to find the best values for such non-zero parameters.  

Under these assumptions, the direct application of {Bayes'} rule over the set of classes and their parameters cannot identify the minimum viable explanation for the experimental data, as, if $\class{a}$ is a sub-model of $\class{b}$, then $\class{b}$ is at least as likely as $\class{a}$ to be the underlying model, once the parameters $\vtheta_{\class{a}}$ and $\vtheta_{\class{b}}$ have been accordingly selected for both classes. In general, there could be a trade-off between the error in explaining the observed data and the complexity of the model selected, which can be tuned through {a} regularization or {a} thresholding parameter.

\subsection{Selecting a minimal model}
\label{subsec:minimal_model_selection}
In this section we discuss the use of regularizing priors and thresholding to select a minimum viable model for explaining the data. We will outline the algorithm capable of extracting the salient features of a model explaining the data, in accordance with a hierarchical view of the class of models. According to the problem this algorithm is being applied to, regularization or thresholding could have different roles and contribute differently to the final outcome. A closer look at the structure of the particular problem at hand is therefore fundamental.

\subsubsection{Regularizing prior}
\label{subsec:prior_regularization}
With a nested set of models, we approach the model selection problem with an ``Occam{'s} razor'' heuristic, favoring the models that can fit the experimental data with the least number of parameters.

In order to penalize models with {a} higher number of non-zero parameters, we substitute the prior $\pi(\vtheta)$ in Eq.~\ref{eq:bayesian} with $\pi(\vtheta) \cdot r(\vtheta)$, where $r(\vtheta)$ is the L$_1$-regularizer. This corresponds to a Laplace distribution
\begin{equation}
    r(\vtheta) = r_1 (\vtheta) := \frac{1}{(2 \sigma)^d} e^{-\frac{\norm{\vtheta}_1}{\sigma}} \; ,
    \label{eq:L1_laplace}
\end{equation}
where the parameter $\sigma$ acts as the regularization scale, $d$ is the size of $\vtheta \in \mathbb{R}^d$, and $\norm{\vtheta}_{1}:= \sum_{i=1}^d |\theta_i|$. With this addition, the ELBO becomes
\begin{align}
    \text{ELBO}' (\boldsymbol{\lambda}) &\coloneqq\ 
    \mathbb{E}_{\boldsymbol{\theta} \sim q_{\boldsymbol{\lambda}}} \Bigg[ 
        \log \pi (\boldsymbol{\theta}) + \log r (\boldsymbol{\theta}) \nonumber \\
    &\qquad 
        + \sum_{t=1}^{M} \log p(y_t \mid x_t, \boldsymbol{\theta}) - \log q_{\boldsymbol{\lambda}} (\boldsymbol{\theta}) 
    \Bigg] \; .
\end{align}
By inserting the explicit expression for $r(\vtheta)$, the ELBO {becomes}:
\begin{equation}
    \text{ELBO}' (\vlambda) := \text{ELBO} (\vlambda) - \mathbb{E}_{\boldsymbol{\theta} \sim q_{\boldsymbol{\lambda}}} \left[ \frac{ \norm{\vtheta}_{1} }{\sigma} \right] \; ,
\end{equation}
where we have neglected a constant term. In this context, Laplace regularization acts as the LASSO in machine learning~\cite{gelmanWeaklyInformativeDefault2008, tibshiraniRegressionShrinkageSelection1996}, and has been previously used, for example, in compressed sensing for quantum tomography~\cite{grossQuantumStateTomography2010} and to reduce the number of CNOT gates in the compilation of quantum circuits~\cite{nemkovEfficientVariationalSynthesis2023}. The L$_1$-regularizer is the most commonly used to find a sparse solution, where those parameters not needed to explain the data will be set to zero {and pruned.} {This regularization is in {the} spirit {of} the generalized Bayesian information criterion. The differences are that the norm does not count directly the number of non-zero parameters, it requires the introduction of an appropriate scale $\sigma$, and {it} can be used online during the training of the Bayesian posterior, unlike GIC.} 

A different choice for the regularization prescribes the use of a Gaussian distribution:
\begin{equation}
    r(\vtheta) = r_2 (\vtheta) := \frac{1}{(2\pi \sigma^2)^{\frac{d}{2}} } 
    \exp \left( -\frac{\norm{\vtheta}_2^2}{2\sigma^2} \right) \; ,
    \label{eq:L2_gaussian}
\end{equation}
where $\norm{\vtheta}_{2}^2:= \sum_{i=1}^d |\theta_i|^2$, which produces the regularized ELBO:
\begin{equation}
    \text{ELBO}' (\vlambda) := \text{ELBO} (\vlambda) - \mathbb{E}_{\boldsymbol{\theta} \sim q_{\boldsymbol{\lambda}}} \left[ \frac{ \norm{\vtheta}_{2}^2 }{2 \sigma^2} \right] \; .
\end{equation}
The Gaussian regularizer corresponds to ridge regression~\cite{kressNumericalAnalysis2012} in machine learning. Instead of favoring a sparse solution, Gaussian regularization finds more balanced solutions, with the tendency to shrink $\vtheta$ towards zero. 

The interpretation of the modified ELBO is that we pay a price each time one of the parameters $\theta_i$ is not exactly zero or close to zero, according to the choice of the norm. This will nudge the training towards models that explain the data correctly, i.e. maximize the ELBO, but also pushes unused parameters to zero.

The choice between L$_1$ or L$_2$-regularization depends on the specific model under consideration. In practice, we observed that the L$_1$ regularization is generally suited {to} sharply exclude features and select between competing models, favoring the least complex ones, while the L$_2$ is useful in scenarios where there is a background signal. In this latter case, we want to better account for this background in the training of the posterior, and postpone the proper model selection to the post-processing, which is now a selection of the most salient features {that} stand out from the background. 

{In this work we have explored the use of both L$_1$ and L$_2$ regularization, and while we have observed L$_2$ to be better in simulation, we have concluded, empirically, that L$_1$ works better on the experimental data. The plots and numerical results reported, for simulated (L$_2$) and experimental data (L$_1$), reflect this choice. We stress that this result might be dependent on the particular dataset and choices of hyperparameters for the inference; a complete analysis of benefits of the two regularizers in model selection goes beyond the scope of this work.}

\subsubsection{Parameters thresholding and confidence level}
\label{subsec:model_confidence_level}
Once the ansatz $q_{\vlambda} (\vtheta)$ for the posterior distribution is trained, the most-likely least-complex model that describes the experimental observations needs to be extracted. According to the selected regularizer, a fraction of estimated parameters will be either exactly zero, or very close to zero. The (post-processing) pruning steps using $q_{\vlambda} (\vtheta)$ are described below.
\begin{enumerate}
    \item $\vtheta \sim q_{\boldsymbol{\lambda}} (\vtheta)$ is sampled and the mask $\theta'_j := \theta_j \cdot \mathds{1}_{\theta_j > \theta_j^\text{th}} (\theta_j)$ is applied. This ensures that when the $j$-th component of the sampled vector is less {than} a threshold $\theta_j^{\text{th}}$, it is exactly set to zero through the multiplication with the indicator function. We define $\vtheta' = (\theta_1', \theta_2', \cdots, \theta_d')$.
    \item The pruning function is then applied to $\vtheta'$, obtaining the minimal model class $\class{c}$ that only has the non-zero identified components, and its associated $\vtheta_{\class{c}}$.
    \item The sampling and the post-processing delineated in 1. and 2. are repeated $Z$ times, and the $\vtheta_{\class{c}}$ are classified according to the number of non-zero parameters, in order to build a set of samples $\mathcal{S}_\class{c} := \lbrace \vtheta_1, \vtheta_2, \cdots \rbrace$ for each $\class{c}\in \mathcal{C}$ observed.
    \item To each class $\class{c}$ we associate a ``pseudo-Bayesian'' probability, defined as $p_\class{c} :=  | \mathcal{S}_\class{c} | /Z$, as the probability for the model $\class{c}$ to be the source of the observed data, i.e. the fraction of the samples of the trained posterior, over all the samples, that have been reduced to the class $\class{c}$.
\end{enumerate}
We identify $\lbrace p_\class{c} \rbrace_{\class{c} \in \mathcal{C}}$ as the probabilities $P(\class{c} \mid  \vx_t, \vy_t)$ on $\mathcal{C}$, while $\mathcal{S}_\class{c}$ {is} a set of samples from the posterior distribution for $\vtheta$ conditioned on $\class{c}$. The class estimator will be the maximum a posteriori estimator (MAP), i.e. $\hat{\class{c}} := \text{argmax} \, p_{\class{c}}$.

\subsection{Uncertainty estimation}
\label{subsec:uncertainty_estimation}
We denote $\lbrace p_{\class{c}} \rbrace_{\class{c} \in \mathcal{C}}$ as pseudo-Bayesian probabilities, as they are not determined from the observed data only, but also depend on the regularization and the thresholding, in short from the model selection algorithm. The role of $p_\class{c}$ is to offer a certainty level for the class $\class{c}$ to be the simplest one explaining the observations. Ideally, we want $p_\class{c}$ to be as concentrated as possible, minimizing uncertainty about the minimum complexity class that fits the data. From the set of samples $\mathcal{S}_\class{c}$, conditioned on the model's class $\class{c}$, we can then evaluate any function of the parameters and their uncertainty after post-processing the samples, see Section~\ref{sec:nanonmr_example}.

\subsection{Nuisance parameters}
\label{subsec:nuisance}
Eq.~\eqref{eq:bornrule} defines the model used in VBI. In some cases, one might not want to treat all parameters in a Bayesian way, for example because their uncertainties are irrelevant {for the purpose of the experimenter} or negligible, but these parameters still need to be included {in the inference} as they impact the observation model. Examples could be some intrinsic properties of a qubit (e.g. its coherence time), or the value of {an} externally applied magnetic or electric field. We call these ``nuisance parameters'' and we label them as $\vphi$. We then write the model as $p(y_t \mid \vtheta, \vphi, x_t)$, with the observation probability depending on the Bayesian parameters $\vtheta$, as well as on the nuisance $\vphi$ and the controls $x_t$. Nuisance parameters can be properties of the probe state, of the measurement, or of the evolution channel. 

The ELBO depends now explicitly on $\vphi$:
\begin{widetext}
    \begin{align}
    \text{ELBO}(\boldsymbol{\lambda}, \vphi) &\coloneqq 
    \mathbb{E}_{\boldsymbol{\theta} \sim q_{\boldsymbol{\lambda}}} \left[ 
        \log \pi(\boldsymbol{\theta})
        \vphantom{\sum_{t=1}^{M}} + \sum_{t=1}^{M} \log p(y_t \mid x_t, \boldsymbol{\theta}, \vphi) - \log q_{\boldsymbol{\lambda}}(\boldsymbol{\theta}) 
    \right]
\end{align}
\end{widetext}
Maximizing the ELBO at the same time with respect to $\vlambda$ and $\vphi$ will perform VBI on $\vtheta$ and maximum expected likelihood estimation on $\vphi$, which follows from the expression of the gradient of the ELBO with respect to $\vphi$, i.e.
\begin{equation}
    \frac{\partial \text{ELBO}(\vtheta, \vphi)}{\partial \vphi} = \sum_{t=1}^M \frac{\partial}{\partial \vphi} \mathbb{E}_{\boldsymbol{\theta} \sim q_{\boldsymbol{\lambda}}} \left[ \log p(y_t \mid x_t, \boldsymbol{\theta}, \vphi) \right] \; .
\end{equation}
In the practical example discussed in Section \ref{sec:nanonmr_example}, various decoherence parameters are treated as nuisance parameters and optimized with maximum likelihood.

\section{Example: quantum sensing of individual nuclear spins}
\label{sec:nanonmr_example}
As an example of the application of our multi-dimensional quantum estimation and model selection protocol, we consider the case of the identification of individual nuclear spins with a single electron spin quantum sensor. This is a problem of practical importance for different quantum technology applications \cite{marcksNuclearSpinEngineering2025}. 

The detection of single nuclear spins is motivated by the desire to push magnetic resonance to the nanoscale \cite{staudacherNuclearMagneticResonance2013, shiSensingAtomicscaleStructure2014, zopesThreedimensionalLocalizationSpectroscopy2018, zopesThreeDimensionalNuclearSpin2018, cujiaTrackingPrecessionSingle2019, cujiaParallelDetectionSpatial2022, herbMultidimensionalSpectroscopyNuclear2024, budakianRoadmapNanoscaleMagnetic2024, duSinglemoleculeScaleMagnetic2024} in order to image single molecules with near-atomic spatial resolution. Quantum sensors based on single electronic spins are promising for achieving this target, as close proximity to the sample enables much higher sensitivity for magnetic dipole signals that decay fast as a function of distance, over a nanoscale sensing volume. 

Additionally, individual nuclear spins are promising for the implementation of long-lived qubits \cite{zhaoSensingSingleRemote2012, wolfowicz29SiNuclear2016, kalbEntanglementDistillationSolidstate2017, bradleyTenQubitSolidStateSpin2019, asaadCoherentElectricalControl2020, abobeihFaulttolerantOperationLogical2022, bradleyRobustQuantumnetworkMemory2022, reinerHighfidelityInitializationControl2024}, as their minimal interaction with the environment leads to long coherence times. These properties make individual nuclear spins attractive building blocks for networked quantum information processing, particularly in architectures where a single electronic spin in a solid-state defect serves as the interface between the local nuclear register and photonic communication channels.

The experiments to identify nuclear spins consist of long pulse sequences to achieve sufficient spectral resolution to discriminate small differences in hyperfine coupling strength. The complexity of the problem and the required fine spectral details lead to long data acquisition times and very large datasets, which are laborious to both acquire and analyze. 

This complexity has stimulated work to find algorithms for automated data processing. Previous approaches \cite{jungDeepLearningEnhanced2021, varona-uriarteAutomaticDetectionNuclear2024} have employed deep learning techniques, framed the problem as pattern recognition or signal-to-image conversion, and demonstrated high performance and speed in identifying nuclear spins from experimental data traces. The drawback of deep learning approaches is, however, that they are tailored to a specific set of measurements, with fixed hyperparameters (e.g. external magnetic field strength) and {to} a specific platform, {requiring an extensive training time before being able to operate.} This is incompatible with adaptivity, as adaptive protocols decide in real time which data is best to acquire and use as input for the evaluation. 

Parallel work by A. Poteshman et al.\cite{poteshman_TransdimensionalHamiltonianModel_2025, poteshman_HighthroughputSpinbathCharacterization_2025} develops a model selection and parameter estimation procedure based on reversible jump Markov chain Monte Carlo, which they deploy {for} the same case study investigated here. The main advantage of the variational approach presented in our manuscript is that it is more suitable {for} exploiting parallelization, with a much shorter training time for the posterior, which can enable the implementation of real-time sequential Bayesian experiment design. We will present a brief comparison between the two approaches in Sec.~\ref{subsec:experimental_results}.

\begin{figure*}[!htbp]
    \centering
\includegraphics[width=1.0\textwidth]{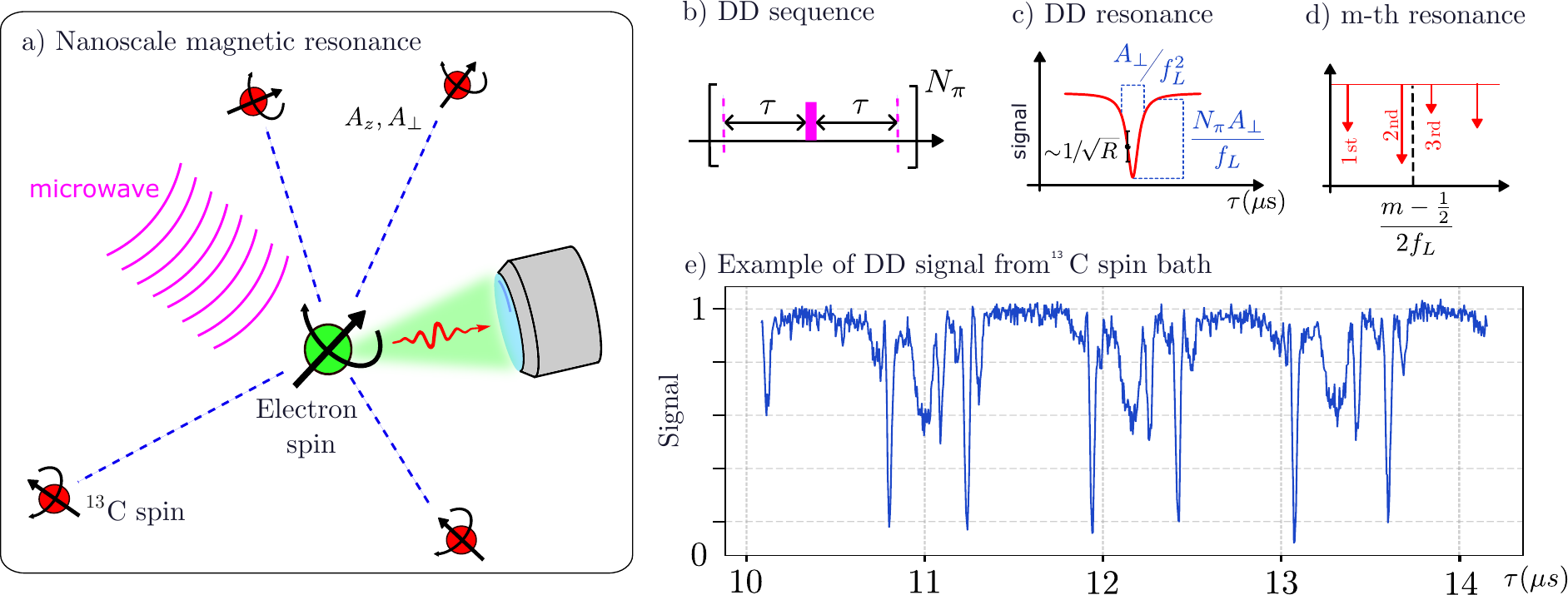}
    \caption{\justifying \textbf{Example of application to nanoscale magnetic resonance experiments.} \textbf{a)} We consider a single electron spin associated with an NV center in diamond, which can be optically initialized and read out. The electron spin is controlled by microwave pulses, and couples to a bath of independent $^{13}$C nuclear spins in the diamond through the hyperfine interaction. \textbf{b)} Measurements consist of dynamical decoupling sequences, comprising $N_{\pi}$ repetitions of the $\left( \pi - \tau - \pi \right)$ unit on the electron spin, initially prepared into an equal superposition state. \textbf{c)} Each nuclear spin, with hyperfine described by the values $\lbrace A_{z, k}; A_{\perp, k} \rbrace$ results in {multiple} resonances with linewidth on the order $\mathcal{O}(A_{\perp} / f_L^2)$, with $f_L$ the Larmor frequency of the spin. \textbf{d)} The signal generated by four spins includes multiple harmonics of index $m$; the shift from the inverse of the Larmor frequency is controlled by $A_z$, i.e. the parallel hyperfine coupling of each spin. \textbf{e)} The outcome signal, upon repeating $R$ experiments consisting of spin initialization, DD sequence, and electron spin readout, is the probability for the electron spin to be in $\ket{0}$ as a function of the inter-pulse delay $\tau$ in the DD sequence. The measurement signal presents a series of dips occurring at values of $\tau$ related to the parallel hyperfine of a given nuclear spin by $\tau_m \approx \frac{m - \frac{1}{2}}{2 f_L + A_{z}}$. The data in the figure {are} a sub-set of the experimental data from M. H. Abobeih et al.\cite{abobeihOnesecondCoherenceSingle2018}.}
    \label{fig:nanonmr_intro}
\end{figure*}

\subsection {Problem statement}
\label{subsec:problem_statement}
We consider the experimental configuration in Fig.~\ref{fig:nanonmr_intro}(a). As a quantum sensor, we consider the $S=1$ electron spin associated with a nitrogen-vacancy (NV) center in diamond. The NV electronic spin is widely used in quantum sensing applications, as it can be optically initialized and readout and features excellent quantum coherence in a wide temperature range, up to room temperature and beyond \cite{doherty_NitrogenvacancyColourCentre_2013, schirhagl_NitrogenVacancyCentersDiamond_2014, barry_SensitivityOptimizationNVdiamond_2020}. 

In a suitable rotating frame and under the secular approximation, the NV single electron spin (indicated by the operator $\mathbf{\hat{S}}$) interacts with surrounding $^{13}$C nuclear spins (operators $\mathbf{\hat{I}_k}, k = 1 ... K$) through the following Hamiltonian \cite{taminiauDetectionControlIndividual2012}:
{
\begin{widetext}
    \begin{equation}
        \begin{split}
            \hat{\mathcal{H}} := D \hat{S}_z^2 & + \gamma_e \mathbf{B} \cdot \mathbf{\hat{S}} + 2 \pi \sum_{k=1}^K \left[ A_{z, k} \hat{S}_z \hat{I}_{z,k} + A_{\perp, k} \hat{S}_z \hat{I}_{x,k}  \right] + \sum_{k=1}^K \gamma_n B_z\hat{I}_{z, k} \; .
        \end{split}
    \label{eq:nanonmr_hamiltonian}
    \end{equation}
\end{widetext}
}
Here $D$ is the NV zero-field splitting ($D = 2 \pi \times 2.87$ GHz), $B_z$ is the external magnetic field applied along the NV axis, $\gamma_n$ is the $^{13}$C gyromagnetic ratio, and $\lbrace \left( A_{z, k}; A_{\perp, k} \right), k = 1 \dots K \rbrace$ are the hyperfine parameters, coupling the NV electron spin with the $^{13}$C nuclear spins.  

This model neglects the dipolar coupling between the nuclear spins, considering them as independent.   {While this approximation is not fully valid, most nuclear-nuclear interactions are small ($<20$ Hz) and can be approximately neglected on the time scales of the experiments considered. } While we use this approximation here for simplicity, the extension of {the} algorithm to include nuclear spin interactions is straight-forward, only computationally more expensive due to the more complex Hamiltonian. 

Our goal is to estimate the hyperfine parameters $\lbrace \left( A_{z, k}; A_{\perp, k} \right), k = 1 \dots K \rbrace$, with the number $K$ of spins {a priori} unknown. We denote $\vA_z := \left( A_{z, 1}, A_{z, 2}, A_{z, 3}, \dots, A_{z, K} \right)$ and $\vA_\perp := \left( A_{\perp, 1}, A_{\perp, 2}, A_{\perp, 3}, \dots, A_{\perp, K} \right)$ the {parallel and orthogonal} hyperfine couplings, respectively, and with
\begin{equation}
    \vA := \left( A_{z, 1}, A_{\perp, 1}, A_{z, 2}, A_{\perp, 2}, \dots, A_{z, K}, A_{\perp, K}\right) \; 
\end{equation}
the full vector of hyperfine couplings. 

We consider here dynamical decoupling measurements performed by preparing the electron spin in an equal superposition $\left(\ket{m_s=0} +\ket{m_s=1} \right)/\sqrt{2}$, and applying a sequence  {$\left( \tau - \pi - \tau \right)$} repeated $N_{\pi}$ times, where “$\pi$’’ represents a $\pi$-pulse on the electron spin, and $\tau$ the inter-pulse delay (Fig.~\ref{fig:nanonmr_intro}(b)). In the frequency domain, this corresponds to a filter \cite{biercukDynamicalDecouplingSequence2011, degenQuantumSensing2017} with central frequency determined by the inter-pulse delay $\tau$, and width inversely proportional to $N_{\pi}$. This is just one of the possible pulse sequences investigated in the literature for the identification of nuclear spins: while more sophisticated sequences have been investigated  \cite{laraouiHighresolutionCorrelationSpectroscopy2013, casanovaRobustDynamicalDecoupling2015, bradleyTenQubitSolidStateSpin2019, zahedian_BlueprintEfficientNuclear_2024}, we take dynamical decoupling as a benchmark example. 

Assuming single-shot electron spin readout \cite{robledo_HighfidelityProjectiveReadout_2011, dinaniBayesianEstimationQuantum2019}, the outcome of a measurement on the electron spin results in a random Bernoulli variable with outcome probability given by \cite{taminiauDetectionControlIndividual2012, abobeihOnesecondCoherenceSingle2018}

\begin{widetext}
    \begin{equation}
        p (y = 0, 1 \mid x = (\tau, N_{\pi}), \vtheta = \vA) := \frac{1}{2} \left( 1 + (-1)^{b} e^{- \left( \frac{2 N_{\pi} \tau}{T_2} \right)^\xi} \prod_{k=1}^K \mathcal{M}  (\vA_k, \tau, N_\pi) \right) \; ,
        \label{eq:nanonmr_model}
    \end{equation}
    with
    \begin{equation}
        \mathcal{M} (\vA_k, \tau, N_\pi) := 1 - m_{k, x}^2 \cdot \frac{(1 - \cos \alpha_k)(1 - \cos \beta)}{1 + \cos \alpha_k \cos \beta - m_{k, z} \sin \alpha_k \sin \beta} \cdot \sin^2 \left(\frac{N_\pi \varphi_k}{2}\right) \; ,
        \label{eq:single_signal}
    \end{equation}
    where {$\widetilde{f}_k \coloneqq \sqrt{(A_{z,k} + f_L)^2 + A_{\perp,k}^2}, \; \widetilde{\omega}_k \coloneqq 2 \pi  \widetilde{f}_k, \; \alpha_k \coloneqq \widetilde{\omega}_k \tau, \; \beta := \omega_L \tau, \; m_{k,z} \coloneqq ( A_{z,k} + f_L)/\widetilde{f}_k, \; m_{k,x} \coloneqq A_{\perp,k}/\widetilde{f}_k, \; \cos\varphi_k \coloneqq \cos\alpha_k \cos\beta - m_{k,z} \sin\alpha_k \sin\beta$, and finally the Larmor frequency of the $^{13}$C is defined as $f_L := \gamma_n B_z$ and $\omega_L := 2 \pi f_L$.}
\end{widetext}
We have written Eq.~\eqref{eq:nanonmr_model} in a way that makes it clear that $b$ is the binomial outcome of the measurement, $x := (\tau, N_\pi)$ is the control, i.e. the parameters that characterize the design of the DD pulse sequence, and $\vtheta=\vA$, i.e. the hyperfine couplings are to be treated in a Bayesian way. Note that with $K$ spins in the estimation the dimension of the space of parameters is $d = 2K$. The time $T_2$ is the characteristic time of the decoherence process, mainly determined by the interactions between the $^{13}$C spins in the bath, which are not {explicitly} included in the model in Eq.~\ref{eq:nanonmr_hamiltonian}. We treat $T_2$ as a nuisance parameter (Section~\ref{subsec:nuisance}).

As depicted in Fig.~\ref{fig:nanonmr_intro}(b)/(c), in the limit of magnetic field {strength} larger than the hyperfine, each nuclear spin results in a series of resonances in the signal, occurring at the inter-pulse delay
\begin{equation}
    \tau_{m, k} \approx \frac{m - \frac{1}{2}}{2 f_L +  A_{z, k}} \; ,
    \label{eq:tau_resonance}
\end{equation}
where $m$ is an integer. Each resonance is Lorentzian in shape, with width $\mathcal{O}\left( \frac{A_{\perp, k}}{f_L^2} \right)$ and depth $\mathcal{O}\left( \frac{N_\pi A_{\perp, k}}{f_L}\right)$. An example of the DD signal as a function of the inter-pulse delay $\tau$ is shown in Fig.~\ref{fig:nanonmr_intro}(d). An intuitive discussion of the Hamiltonian in Eq.~\ref{eq:nanonmr_hamiltonian} and the corresponding DD signal is presented in Appendix~\ref{sec:intuitive}.

The model (Eq.~\eqref{eq:nanonmr_model}) manifests two symmetries. Given a set of hyperfine couplings $\vA$ we can permute the labels of the spins and obtain the same signal. Furthermore, the model is symmetric for $A_{\perp, k} \rightarrow - A_{\perp, k}$, which gives yet another degeneracy. In the estimation, we will fix  $A_{\perp, k} \ge 0$. The development of symmetry-preserving ansatzes is an open problem in VBI, which we briefly discuss in Section~\ref{sec:outlook}. To conclude this section, we observe that the spatial location of a nuclear spin is not unequivocally determined by knowledge of the $(A_z, A_{\perp})$ couplings. The hyperfine couplings are however relevant parameters needed to address the spins individually and implement quantum gates \cite{taminiauDetectionControlIndividual2012}.

\subsection{Outcome model in VBI}
\label{subsec:outcome_model}
We define a single experiment to comprise $R=10^3$ repetitions of the process of initialization of the NV center, application of the $N_{\pi}$-pulses DD sequence, and measurement. We define the outcome of an experiment as
\begin{equation}
    y := \frac{1}{R} \sum_{r=1}^R b_r + \delta \; , 
\end{equation}
where $\delta \sim \mathcal{N}(0, \eta_0^2)$ is the measurement noise in {addition} to the shot noise, and we have:
\begin{equation}
    \lim_{R\rightarrow \infty} \mathbb{E} [y] = p := p ( 1 \mid (\tau, N_{\pi}), \vA)
\end{equation}
The variance (noise) on the measurement $y$ is the sum of the shot noise plus the Gaussian error, which models for us all other sources of errors in the experiment ($\eta_0 := 10^{-2}$ {for the data generated in simulation}), i.e.
\begin{equation}
    \var{y} := \frac{ p (1 - p)}{R} + \eta_0^2 \; .
    \label{eq:noise_variance}
\end{equation}
In the estimation we choose to model the outcomes as Gaussian variables, sampled from the distribution
\begin{equation}
    y \sim \mathcal{N} \left( p, { \frac{p(1-p)}{R} + \eta^2} \right) \; ,
    \label{eq:gaussian_noise}
\end{equation}
where $\eta$ (initialized at $\eta_0=10^{-2}$), and $T_2^{-1}$ (initialized at zero), are the nuisance parameters in the estimation, i.e. $\vphi := (T_2^{-1}, \eta)$, and are therefore estimated with expected maximum likelihood, as explained in Sec.~\ref{subsec:nuisance}. We stress that, while the generation of the simulation dataset involves sampling  {from a binomial distribution}, the model we use in VBI treats $y$ as a Gaussian variable. We have used the same model for processing both simulated and experimental data. Notice also that the probability distribution for $y$ has a complicated dependence on the hyperfine couplings $\vA$ {through} the probability $p$. In the following sections we indicate with $\mathcal{G}(y \mid x, \vA, \vphi)$ the density function of the Gaussian in Eq.~\eqref{eq:gaussian_noise}, which plays the role of the model for the outcome probability in nano-NMR, i.e. $p(y|x, \vtheta)$ in Eq.~\eqref{eq:bornrule}.

\subsection {Algorithm deployment}
\label{subsec:deployment}
Here we illustrate the application of the multi-parameter quantum estimation and model selection procedure described in Sections~\ref{sec:bayesian_learning} and~\ref{sec:model_learning} to the nanoscale magnetic resonance problem. 

The expression for the ELBO in Eq.~\eqref{eq:elbo_definition} is evaluated numerically by approximating the expectation value over the ansatz with a Monte Carlo approach, i.e. we compute $\mathcal{B}$ extractions of hyperfine couplings from the current ansatz $q_{\vlambda} (\vA)$ to approximate the ELBO as the following average:
\begin{widetext}
    \begin{equation}
    \begin{split}
        \text{ELBO} := \frac{1}{\mathcal{B}} \sum_{j=1}^\mathcal{B} \biggl[ & \log \pi(\vA) + \log r(\vA) + \sum_{t=1}^{M_B} \log \mathcal{G}(y_t \mid x_t, \vA, \vphi)  - \log q_{\vlambda}(\vA) \biggr] \; .
    \end{split}
    \end{equation}
    \label{eq:elbo}
\end{widetext}
Throughout the paper we use an uninformative, unnormalized prior $\pi(\vA) \propto 1$. We always use $M_B = M$, i.e. we process the whole dataset in each computation of the ELBO, but it is possible to select smaller mini-batches to ease the computational requirements of the inference process. For the processing of nano-NMR we use a normalizing flow as the ansatz, constituted by five layers of neural autoregressive layers~\cite{huangNeuralAutoregressiveFlows2018} followed by one last affine layer. See Sec.~\ref{sec:bayesian_learning} for a definition of the normalizing flow.

\subsection{Posterior post-processing}
\label{subsec:postprocessing_nanonmr}
In Sec.~\ref{subsec:model_confidence_level} we have outlined the working principles of the post-processing of the trained posterior. This is used to prune unnecessary parameters and obtain the minimal model that explains the data. For the particular case of nano-NMR, a lower threshold $A_\perp^{\text{th}}$ is imposed, while there is no lower threshold on $A_z$, although we choose to process and visualize only samples within $A_z < A_z^{\text{th}} \simeq 80 \,\kHz$. When the estimation is carried out with an ansatz for the posterior containing  {more spins than required to accurately describe the experimental data}, the spins in excess are pushed below the threshold by the maximization of the ELBO, and are finally cut by the thresholding. Also spins that are in the signal, but are too weak to be above threshold, are cut. The threshold therefore represents the cut-off of a blind zone, beyond which the coupling of the spins with the NV center is too weak for the spins to be individually detected {from the dataset considered}.

The thresholding is performed only on $A_\perp$ and not on $A_z$ because the orthogonal component of the hyperfine coupling is the parameter controlling the amplitude and width of the dips, i.e. the main features determining whether a dip is detectable within the noise level.

With an ansatz containing $K$ spins, excluding the class with zero spins, the thresholding procedure described in Sec.~\ref{subsec:model_confidence_level} leaves us with $K$ sets $\mathcal{S}_1, \mathcal{S}_2, \cdots, \mathcal{S}_K$, which contain all the samples classified according to the number of spins, i.e. 
\begin{widetext}
    \begin{equation}
        \begin{array}{ccc}
            \mathcal{S}_1 := \left\{
                \begin{array}{c}
                    (A_z^1, A_\perp^1)_{1} \; , \\
                    (A_z^1, A_\perp^1)_{2} \; ,  \\
                    (A_z^1, A_\perp^1)_{3} \; ,\\
                    \vdots \\
                    (A_z^1, A_\perp^1)_{|\mathcal{S}_1|} \; .
                \end{array}
            \right\} ,
            &
            \mathcal{S}_2 := 
            \left\{
                \begin{array}{c}
                    (A_z^1, A_\perp^1, A_z^2, A_\perp^2)_1 \; ,\\
                    (A_z^1, A_\perp^1, A_z^2, A_\perp^2)_2 \; , \\
                    (A_z^1, A_\perp^1, A_z^2, A_\perp^2)_3 \; , \\
                    \vdots \\
                    (A_z^1, A_\perp^1, A_z^2, A_\perp^2)_{|\mathcal{S}_2|} \; .
                \end{array}
            \right\}
            &  , \dots \; \mathcal{S}_n :=
            \left\{
                \begin{array}{c}
                    (A_z^1, A_\perp^1, A_z^2, A_\perp^2, \cdots, A_z^n, A_\perp^n)_1 \; , \\
                    (A_z^1, A_\perp^1, A_z^2, A_\perp^2, \cdots, A_z^n, A_\perp^n)_2 \; , \\
                    (A_z^1, A_\perp^1, A_z^2, A_\perp^2, \cdots, A_z^n, A_\perp^n)_3 \; , \\
                    \vdots \\
                    (A_z^1, A_\perp^1, A_z^2, A_\perp^2, \cdots, A_z^n, A_\perp^n)_{|\mathcal{S}_n|} \; .
                \end{array}
            \right\} \; , \cdots
        \end{array}
    \end{equation}
\end{widetext}
So far we have been following the general framework laid down in Sec.~\ref{sec:model_learning}. The rest of this section is specific to the determination of the number of spins generating the signal (model selection) and the determination of the hyperfine couplings (parameter estimation). For each dataset $\mathcal{S}_n$ containing $|\mathcal{S}_n|$ samples, we break the correlations between the spins in each single sample from the posterior and we build an ensemble of single spins
\begin{equation}
    \widetilde{\mathcal{S}}_n := \lbrace (A_{z, j}, A_{\perp, j})\rbrace_{j=1}^{n |\mathcal{S}_n|} \; ,
\end{equation}
for $n \le K$. These are now samples from a 2D probability distribution, which is a marginalized version of the full posterior distribution. Although this process discards much information, it is a necessary step to obtain a condensed version of the posterior for two-dimensional plotting and clustering. We then apply the K-means{s} clustering algorithm to this dataset, with an optimal number of clusters obtained by testing the inertia for each number of clusters between $1$ and $n$. This procedure leaves us with a partition of $\widetilde{\mathcal{S}}_n$ in sets of points belonging to different clusters:
\begin{equation}
    \widetilde{\mathcal{S}}_n = \bigcup_{j=1}^{L_n} K_{n, j} \; ,
\end{equation}

\noindent with $L_n$ being the total number of clusters. For each cluster of points $K_{n, j}$ we then compute three quantities: 

1) the average value of the 2D vector of couplings, i.e.
\begin{equation}
    \vmu_{n, j} := \frac{1}{|K_{n, j}|} \sum_{(A_z, A_\perp) \in K_{n, j}} (A_z, A_\perp) \; .
\end{equation}

2) the covariance matrix of the cluster, i.e.
\begin{equation}
    \Sigma_{n, j} := \frac{1}{|K_{n, j}|}   \sum_{(A_z, A_\perp) \in K_{n, j}}  
    \begin{pmatrix} A_z \\ A_\perp \end{pmatrix}
    \! 
    \begin{pmatrix} A_z & A_\perp \end{pmatrix}
    -  \vmu_{n, j}^T \vmu_{n, j} \; .
\end{equation}

3) the “weight” of the cluster, i.e. 
\begin{equation}
    \#S_{n, j} := \frac{|K_{n, j}|}{|\mathcal{S}_n|} \; .
\end{equation}
The weight of the cluster gives us information on the number of spins in each cluster, and it can be a fractional number. The weights of all clusters are normalized to the number of spins $n$, i.e. $\sum_{j=1}^{L_n} \#S_{n, j} = n$. 

To understand its physical meaning, consider an example case where all samples in $\mathcal{S}_n$ contain a spin at a certain position in the coupling space $(A_z, A_\perp)$. Since the couplings are permutationally invariant, the same spin can have a different label in different samples from the posterior. When breaking the correlations to build $\widetilde{S}_n$, exactly $|\mathcal{S}_n|$ pairs $(A_z, A_\perp)$ refer to the same physical spin. If these are clustered together in $K_{n, j}$, because they are well-separated from all the other spins, a weight of $\#S_{n, j} = 1$ is assigned to the cluster according to the definition. This indicates a single spin at that position in the hyperfine couplings space. If, for example, a posterior distribution for a single {spin} produces samples that cluster with equal probability around two well-separated positions A and B, then each of these clusters will carry a weight $\#S = 0.5$, meaning that the posterior, given the observed data, predicts as equally likely configurations with a spin at position A or a spin at position B.

All intermediate configurations are possible, with an arbitrary number of spins, not exceeding $K$, distributed among any number of clusters. \subsection{Defining performance metrics}
\label{subsec:ml_metrics}
We benchmark the performance of the model selection algorithm by computing the standard machine learning (ML) metrics of precision, recall, F$_1$ score, and the mean absolute error. These are widely used in the ML literature and are based on counting the true positives (TP), false positives (FP), and false negatives (FN). Broadly speaking, we look at clusters obtained in the post-processing of the posterior, and if an appropriate number of spins in the ground truth is close to the cluster, we consider these spins to be counted among the true positives. If the algorithm signals a cluster, but no spins are to be found nearby, we classify this cluster as a false positive. Finally, if no cluster is in the vicinity of a spin in the ground truth, this spin is counted {as} one of the false negatives. The ML figures of {merit} are defined as {follows}:
\begin{equation}
    \begin{split}
    & \text{Precision} := \frac{\text{TP}}{\text{TP}+\text{FP}}, \\
    & \text{Recall} := \frac{\text{TP}}{\text{TP}+\text{FN}}, \\
    & \text{F}_1 := \frac{2 \cdot \text{Precision} \cdot \text{Recall}}{\text{Precision} + \text{Recall}}.
    \end{split}
    \label{eq:recall_precision}
\end{equation}

Precision is the fraction of correctly identified spins over the total number of spins in the ground truth, while recall is the fraction of correctly identified spins over the total number of identified spins. We arbitrarily take the F$_1$ score as the harmonic mean of precision and recall: in general, they should be combined in a figure of merit that acknowledges the costs associated {with} a false positive and a false negative detection based on the specific application under consideration.

The figures of merit are computed as follows. We fix the number $n$ of spins in the conditioned posterior, and for each cluster, indexed by $j$, we compute the Mahalanobis distance \cite{mahalanobis_GeneralizedDistanceStatistics_1936} from each of the spins, indexed by $k$, i.e.
\begin{equation}
    d_{j, k} := \sqrt{(\vmu_{n, j} - \vA_k)^{T} \Sigma_{n, j}^{-1} (\vmu_{n, j} - \vA_k)} \; .
\end{equation}
From this distance, we compute the ML metrics as follows:
\begin{itemize}
    \item We identify all spins for which $d_{j, k} \le t $, with $t \sim \mathcal{O}(1)$ being an arbitrarily chosen threshold value. These are flagged as {true positives}, i.e. they {have} been correctly identified in the cluster. We indicate the number of true positives in a cluster with TP$_{j}$. In this manuscript, we have chosen $t=4$. Assuming that the cluster is Gaussian, this means a probability of $13.5\%$ for a point belonging to the cluster to be flagged as a false negative.
    \item The weight of the cluster $\#S_{n, j}$ is rounded, and if the difference between $\lfloor \#S_{n, j} \rceil$ and the number of true spins identified through the Mahalanobis distance is positive, these spins are considered false positives. The number of false positives in the cluster is
    \begin{equation}
        \text{FP}_{j} := \lfloor \#S_{n, j} \rceil - \text{TP}_j \; .
    \end{equation}
    \item The total number of false positives is $\sum_{j} \text{FP}_{j}$, assuming $\text{FP}_{j} > 0$. The total number of true positives is the number of spins that have been flagged as such, and the number of {false negatives} is the number of true spins that have not been flagged.
\end{itemize}
A cluster can contain multiple spins, and we do not trigger any event in the case {where} more spins are close to the cluster than the weight of the cluster itself. Underestimating the number of spins in a cluster is not penalized, as long as a cluster is identified that is close to the spins. This is in line with the idea of using VBI for a fast exploration of the environment with the goal of obtaining a rough representation of the environment to be used in the adaptive step.

Mathematically, this difference can be seen in the fact that FP$_j > 0$ negatively impacts the precision, while FP$_j < 0$ does not contribute to the false negative{s}, and therefore to the recall, as long as there is a cluster {in} the vicinity of all the spins in the ground truth.

\subsection{Simulation results}
\label{subsec:simulation_results}
In this section, we report the results of simulations carried out for a magnetic field $B = 403 \, \text{G}$, a number $N_{\pi} = 32$ of $\pi$ pulses in the DD sequence, $R=10^3$, batch size $\mathcal{B}=128$, and $I=8192$ training steps. We impose L$_2$ regularization, with $\sigma = 10^{-2}$ and a relatively high value $A_\perp^{\text{th}} = 8 \,\kHz$ for the threshold on the orthogonal hyperfine components. 

We simulate the signal generated by DD experiments on a single electron spin by environments comprising between $6$ and $16$ spins, while the ansatz for the posterior is always trained on $K=20$ spins. Model selection consists of the identification of the exact number of spins generating the signal. The extra spins in the posterior, with couplings close to zero, are pruned by the thresholding procedure.

We have carried out $50$ simulated estimations in total, i.e. one for each number of spins $6 \le n \le 16$ and for each number of measured points in the signal $M=256, 512, 1024, 2048, 4096$. In Table~\ref{table:tau_times} we report the windows used for the inter-pulse time delay in DD, corresponding to each number of measurement points $M$ in the dataset. A data point is taken every $10$\,ns, which is the temporal resolution of the instruments.
\begin{table}[h]
    \centering
    \caption{Range of $\tau$ values for different $M$ used in simulation.}
    \label{table:tau_times}
    \begin{tabular}{c|ccccc}
    $M$ & 256 & 512 & 1024 & 2048 & 4096 \\
    \hline
    $\tau_{\min}$ ($\mu$s) & 6.0 & 6.0 & 6.0 & 6.0 & 6.0 \\
    $\tau_{\max}$ ($\mu$s) & 7.25 & 8.5 & 11.0 & 16.0 & 26.0 \\
    \bottomrule
    \end{tabular}
\end{table}

\begin{figure*}[!htbp]
    \centering
\includegraphics[width=1.0\textwidth]{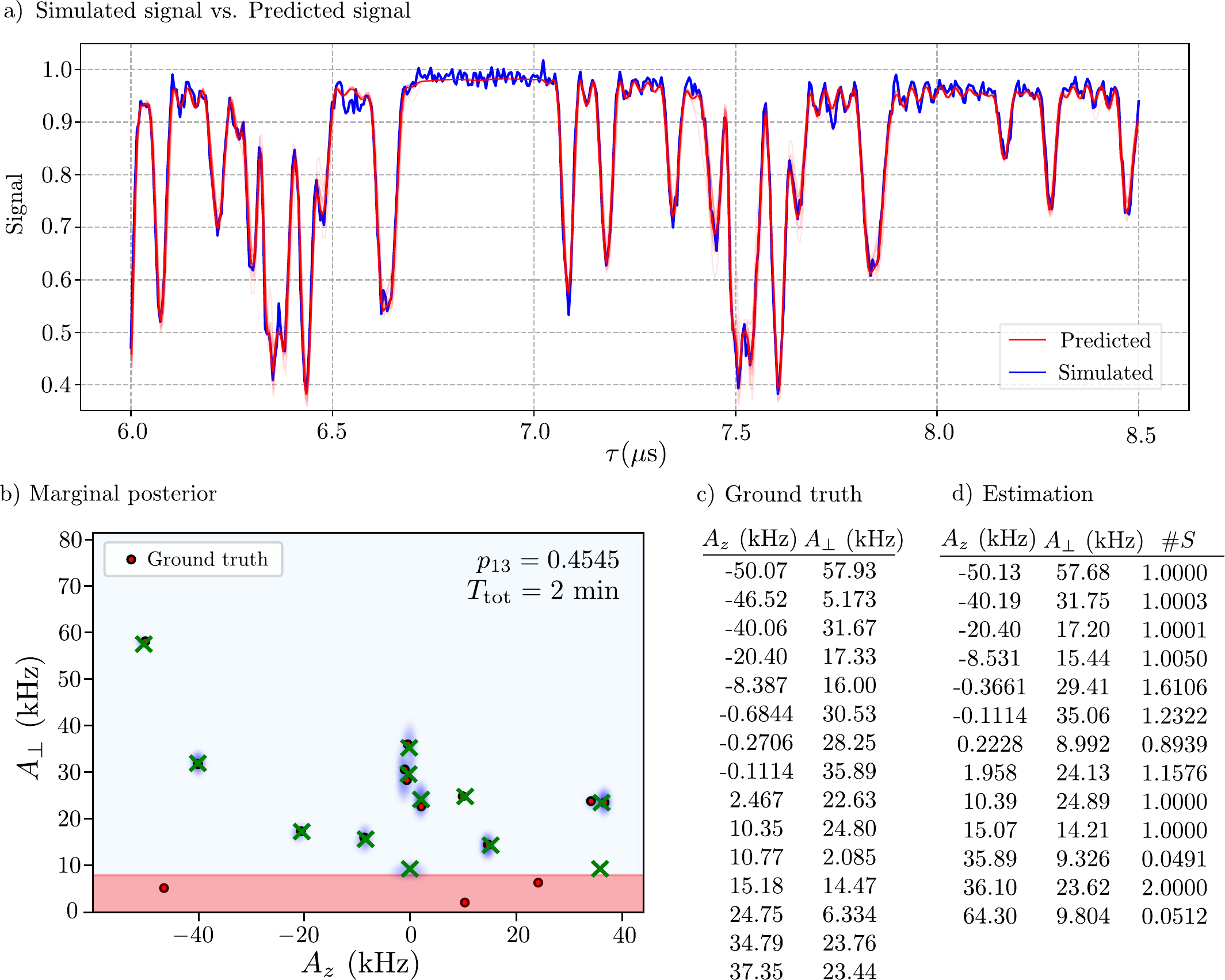}
    \caption{\justifying \textbf{Example of application to a simulated bath of $15$ spins, using $M=512$ values of the inter-pulse delay $\tau$ in the DD measurement.} \textbf{a)} Dynamical decoupling signal (Eq.~\ref{eq:nanonmr_model}) as a function of the inter-pulse delay $\tau$ with $\tau \in [6, 8.5]\,\mu$s.  {We compare the simulated data (blue curve) to the signal calculated from the VBI analysis output parameters (red curve).} The uncertainty in the prediction is represented by the light red lines around the measurement signal, being approximately one standard deviation from the average prediction. \textbf{b)} Marginal posterior distribution trained on the observed signal, conditioned on the presence of $13$ spins (which has been found by the algorithm to be the most likely number of clusters). Three of the spins fall within the blind zone, represented by the red stripe, and cannot be individually detected based on the simulated data. The simulated data includes shot noise and additional measurement noise, as reported in Eq.~\ref{eq:noise_variance} and Eq.~\eqref{eq:gaussian_noise}. The red dots represent the true values of the couplings, the green crosses are the centers of the clusters, and the blue shaded regions indicate the regions where the posterior probability concentrates. \textbf{c)} List of the hyperfine values for the ground truth spins (used to generate the simulated DD signal in (a)). \textbf{d)} List of the hyperfine values for the identified spins, given as the centers of the clusters in the trained posterior, with the corresponding weights. The spins not used in the explanation of the signal (i.e. with couplings close to zero) are pushed into the blind zone by L$_2$ regularization and pruned by the thresholding. The confidence that the environment comprises 13 spins is $p_{13} = 0.4545$, the total signal acquisition time ($T_{\text{tot}}$) is $2$ minutes.}
    \label{fig:simulated_512_meas}
\end{figure*}

\begin{figure*}[!htbp]
    \centering
\includegraphics[width=1.0\textwidth]{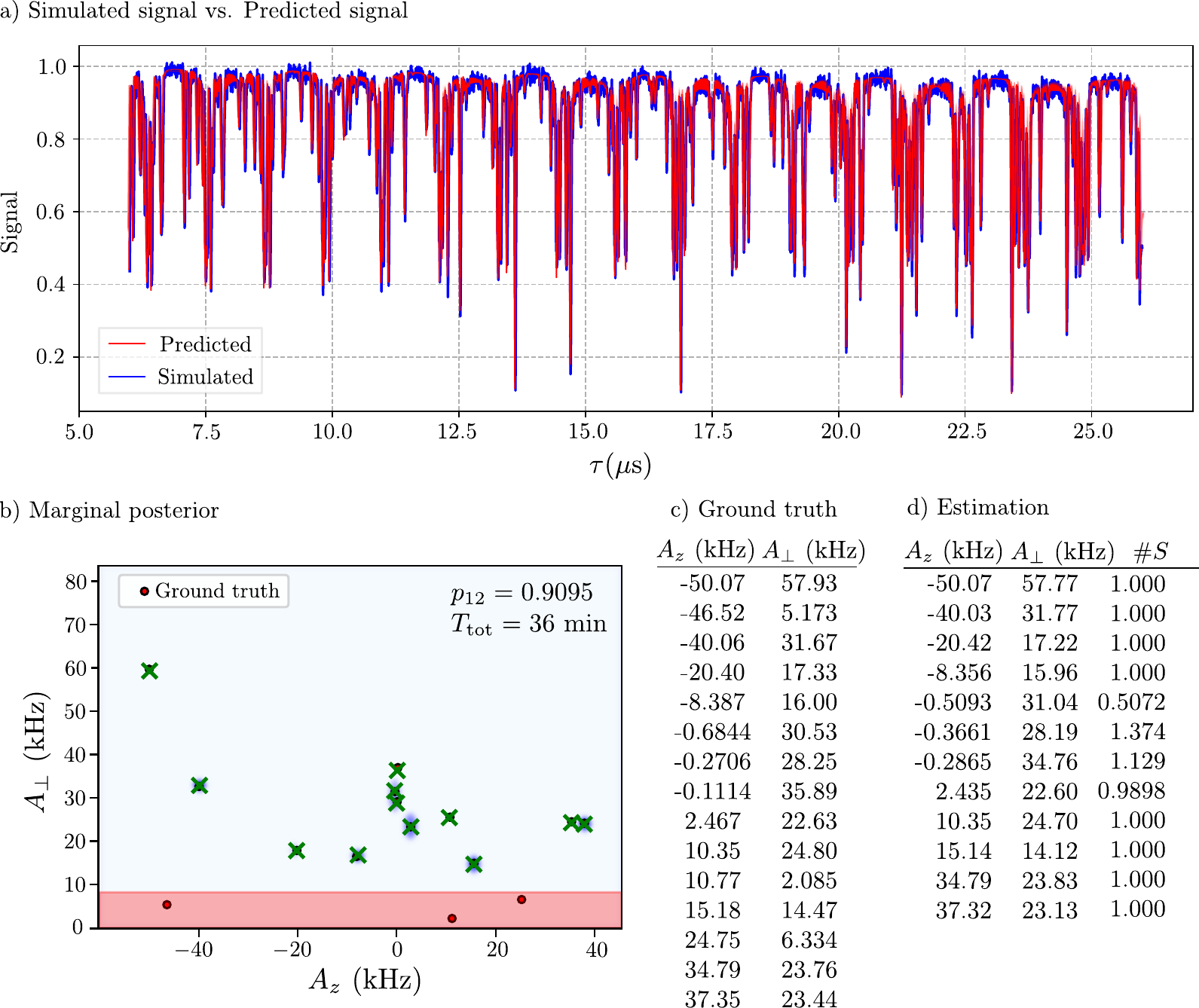}
    \caption{\justifying \textbf{Example of application to a simulated bath of $15$ spins, using $M=4096$ values of the inter-pulse delay $\tau$ in the DD measurement.} This figure reports the results of the algorithm for the same situation as Fig.~\ref{fig:simulated_512_meas}, only with more data, corresponding to $M=4096$ values of $\tau$. The legend for all the elements of the plot is the same as in Fig.~\ref{fig:simulated_512_meas}. By utilizing more data, the algorithm identifies with great confidence the presence of $13$ spins in the environment outside of the blind zone ($p_{13} = 90\%$). Here $T_{\text{tot}} = 36$ minutes.}
\label{fig:simulated_4092_meas}
\end{figure*}

In Fig.~\ref{fig:simulated_512_meas} and Fig.~\ref{fig:simulated_4092_meas}, we report the outcome of our model selection algorithms for, respectively, $M=512$ and $M=4096$ measurement steps. The measurement points are uniformly distributed in the interval $(\tau_{\min}, \tau_{\max})$. Subplot (a) compares the simulated signal (blue line) with the algorithm fit (red line), showing excellent match in all cases. 

Subplot (b) shows the 2D marginalized posterior generated by the VBI procedure (blue), with the ground truth hyperfine values marked as red circles. We have selected the conditional probability corresponding to the most likely number of spins, according to $p_n$, as defined in Sec.~\ref{subsec:model_confidence_level}. The probability density represented by the blue regions indicates the uncertainty around each spin, as estimated by the algorithm. The red stripe marks the threshold on the  {orthogonal} hyperfine coupling. Note that, in plotting only the marginal of the posterior, we are discarding the information about the correlations between the spins (Section~\ref{subsec:postprocessing_nanonmr}). This information can however be useful in the computation of the error for quantities depending on multiple spins, or in providing more accurate confidence levels to inform adaptive measurements.

Subplots (c) and (d) list, respectively, the ground truth hyperfine values for the $n=15$ spins and the estimated values. As explained above, the number  of spins in each cluster is not  necessarily an integer, and it can be interpreted in a probabilistic sense as the average number of spins at that position according to the posterior.

The total measurement duration is computed as $T_{\text{tot}} = \sum_{j=1}^M R \left( 2 N_\pi \tau_j + T_{\text{oh}} \right)$, with $T_{\text{oh}}$ being an overhead time that takes into account $10\,\mu\text{s}$ for the initialization and readout of the NV center for each DD sequence, and $37$ ns for each $\pi$-pulse \cite{abobeihOnesecondCoherenceSingle2018}. We ignore the non-deterministic overhead introduced by NV charge state instabilities, spectral diffusion, and {(optional)} measurement-based initialization of the NV's own $^{14}$N nuclear spin.

\begin{figure*}[!htbp]
    \centering
\includegraphics[width=\textwidth]{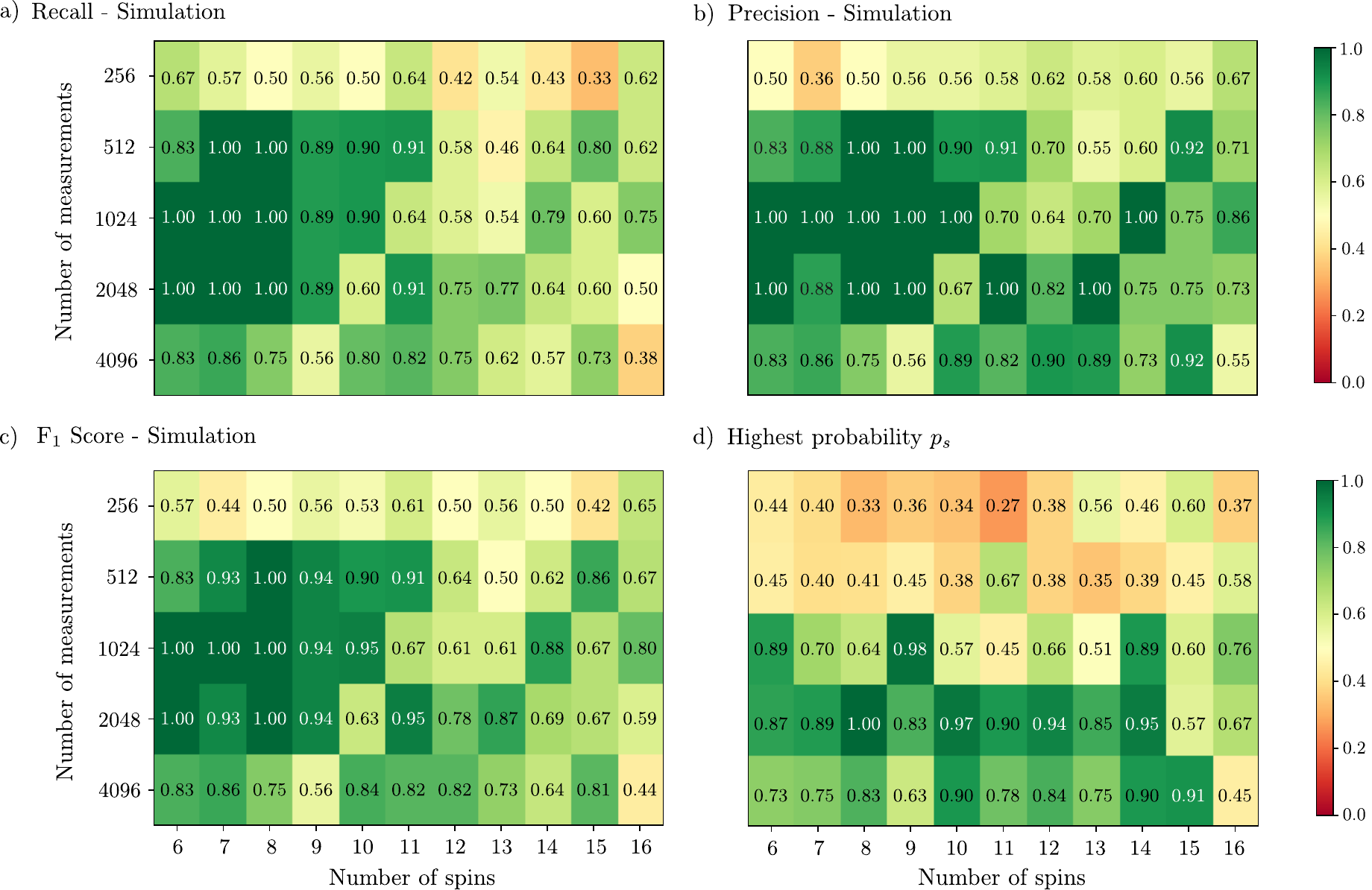}
    \caption{\justifying \textbf{Machine learning metrics for the deployment of the algorithm on simulated data.} We report the machine learning figures of merit for the model selection and estimation of hyperfine couplings in nano-NMR. Samples from the trained posterior are post-processed to extract clusters from the identified spins, as described in Sec.~\ref{subsec:model_confidence_level} and Sec.~\ref{subsec:postprocessing_nanonmr}. The centers and covariance matrices of these clusters are used to identify the false positive and false negative spins, from which the recall, precision, and F$_1$ score are computed, as defined in Eq.~\eqref{eq:recall_precision}. The values of all these figures of merit are in $[0, 1]$, with one corresponding to maximum performance. Sub-plots \textbf{a)}, \textbf{b)}, and \textbf{c)} are tables reporting recall, precision, and F$_1$ score respectively, as a function of the number of spins and number of measurements. Recall represents the impact of the false negatives, precision the impact of the false positives, and the F$_1$ score is the harmonic mean of precision and recall. We observe that the performance using $M=4096$ measurements degrades compared to $M=2048$;  {this happens because the posterior has not been trained to convergence}. \textbf{d)} Probability $p_s$ of the selected number of spins, as defined in Sec.~\ref{subsec:model_confidence_level}. This parameter is a proxy for the confidence level in the identification of the number of spins.}
    \label{fig:complete_metrics}
\end{figure*}

In Fig.~\ref{fig:complete_metrics} we report the ML metrics (recall, precision, and F$_1$-score) obtained in simulation for multiple environments and multiple numbers of measurements.

Finally, we report the results for the average error (in kHz) for the simulated estimation of $A_z$ and of $A_{\perp}$, in Table~\ref{table:error_parallel} and Table~\ref{table:error_orthogonal}, respectively.

\begin{table}[h]
\centering
\caption{\justifying Average error per spin for the parallel component of the hyperfine coupling $A_z$, shown in kHz (high-precision values), as evaluated in simulation.}
\begin{tabular}{c|ccccc}
    &  \multicolumn{5}{c}{{Number of Measurements}} \\
Num. Spins & {256} & {512} & {1024} & {2048} & {4096} \\ \hline
6  & 0.732 & 0.095 & 0.064 & 0.035 & 0.116 \\
7  & 1.035 & 0.605 & 0.076 & 0.100 & 0.089 \\
8  & 0.255 & 0.191 & 0.115 & 0.088 & 0.140 \\
9  & 0.140 & 0.119 & 0.089 & 0.123 & 0.088 \\
10 & 0.812 & 0.102 & 0.107 & 0.060 & 0.056 \\
11 & 0.318 & 0.159 & 0.430 & 0.089 & 0.076 \\
12 & 0.573 & 0.653 & 0.477 & 0.134 & 0.078 \\
13 & 0.175 & 0.684 & 0.605 & 0.107 & 0.121 \\
14 & 0.116 & 0.286 & 0.091 & 0.038 & 0.255 \\
15 & 0.589 & 0.334 & 0.032 & 0.462 & 0.045 \\
16 & 1.210 & 0.557 & 0.140 & 0.557 & 0.525 \\
\end{tabular}
\label{table:error_parallel}
\end{table}

\begin{table}[h]
\centering
\caption{\justifying Average error per spin for the orthogonal component of the hyperfine coupling $A_\perp$, shown in kHz (high-precision values), as evaluated in simulation.}
\begin{tabular}{c|ccccc}
    & \multicolumn{5}{c}{{Number of Measurements}} \\
Num. Spins & {256} & {512} & {1024} & {2048} & {4096} \\ \hline
6  & 1.45  & 0.350 & 0.716 & 0.239 & 0.382 \\
7  & 1.19  & 2.55  & 0.477 & 0.589 & 0.255 \\
8  & 2.39  & 1.19  & 0.732 & 0.621 & 0.350 \\
9  & 1.59  & 0.509 & 0.477 & 0.589 & 0.557 \\
10 & 3.18  & 0.493 & 0.350 & 0.366 & 0.509 \\
11 & 1.53  & 0.557 & 0.334 & 0.366 & 0.350 \\
12 & 1.59  & 1.34  & 1.13  & 0.302 & 0.286 \\
13 & 1.75  & 1.37  & 0.541 & 0.334 & 0.430 \\
14 & 1.24  & 0.955 & 0.398 & 0.255 & 0.828 \\
15 & 0.732 & 0.525 & 0.207 & 1.54  & 0.175 \\
16 & 3.34  & 0.923 & 0.684 & 1.75  & 1.75  \\
\end{tabular}
\label{table:error_orthogonal}
\end{table}

In both tables, the configuration of spins from which the simulated signal is created is representative of a distribution of spins with relatively large hyperfine couplings only. Real configurations typically include a large number of spins further away from the central electron spin, with small hyperfine couplings that produce a background signal. Such spins are not simulated here as they produce a weak background signal and cannot be accurately estimated. While we have neglected the spins further from the NV center here for simplicity, we will see in the next section how VBI-based model selection performs well also under realistic circumstances.

The training times for the algorithm, reported in Table~\ref{table:training_time}, barely depends on the true number of spins in the environment, while exhibiting a stronger dependence on the number of spins in the ansatz and number of measurements. In Table~\ref{table:training_time} we report also $T_{\text{tot}}$, i.e. the total time required for the data acquisition, which, by construction, does not depend on the number of spins in the environment.

\begin{table}[h]
\centering
\caption{\justifying Training time of the VBI ansatz (in minutes) for various numbers of measurements, together with the time $T_{\text{tot}}$ required for acquiring the corresponding signal in the experiment (in minutes) and the training time for the MCMC. The training of the VBI is performed on a GPU, while the MCMC is done on a CPU, see Sec.~\ref{subsec:methods} for more information on the hardware.}
\begin{tabular}{c|ccccc}
    & \multicolumn{5}{c}{Number of Measurements} \\ 
    & {256} & {512} & {1024} & {2048} & {4096} \\ 
    \hline
    $T_{\text{tot}}$  & 1 & 2 & 5 & 16 & 36 \\
    Training time VBI & 5 & 5 & 6 & 8 & 12 \\
    Training time MCMC & 15 & 22 & 35 & 108 & 144 \\
\end{tabular}
\label{table:training_time}
\end{table}

\subsection{Experimental results}
\label{subsec:experimental_results}
We deploy our algorithm on experimental data from M. H. Abobeih et al.~\cite{abobeihOnesecondCoherenceSingle2018}, consisting of dynamical decoupling measurements on a single NV in diamond with natural 1.1\% abundance of $^{13}$C isotopes, pre-selected only for the absence of  $^{13}$C spins with hyperfine coupling larger than 500 kHz. Data was collected at cryogenic temperature, exploiting single-shot electron spin readout through spin-selective resonant excitation, with average fidelity $95$\% \cite{abobeihOnesecondCoherenceSingle2018}. 

The experimental signal has been processed with an ansatz for the posterior containing $K=40$ spins, i.e. $80$ spin parameters. The batch size was set to $\mathcal{B}=256$, and we used the L$_1$ regularization with $\sigma=10^{-2}$, and $R=500$. The threshold on the hyperfine coupling has been decreased to $A_\perp^{\text{th}} = 3 \,\kHz$ compared to the simulations in the previous section. The other parameters are identical to the simulated estimation discussed in Section~\ref{subsec:simulation_results}.

\begin{figure*}[!htbp]
    \centering
\includegraphics[width=1.0\textwidth]{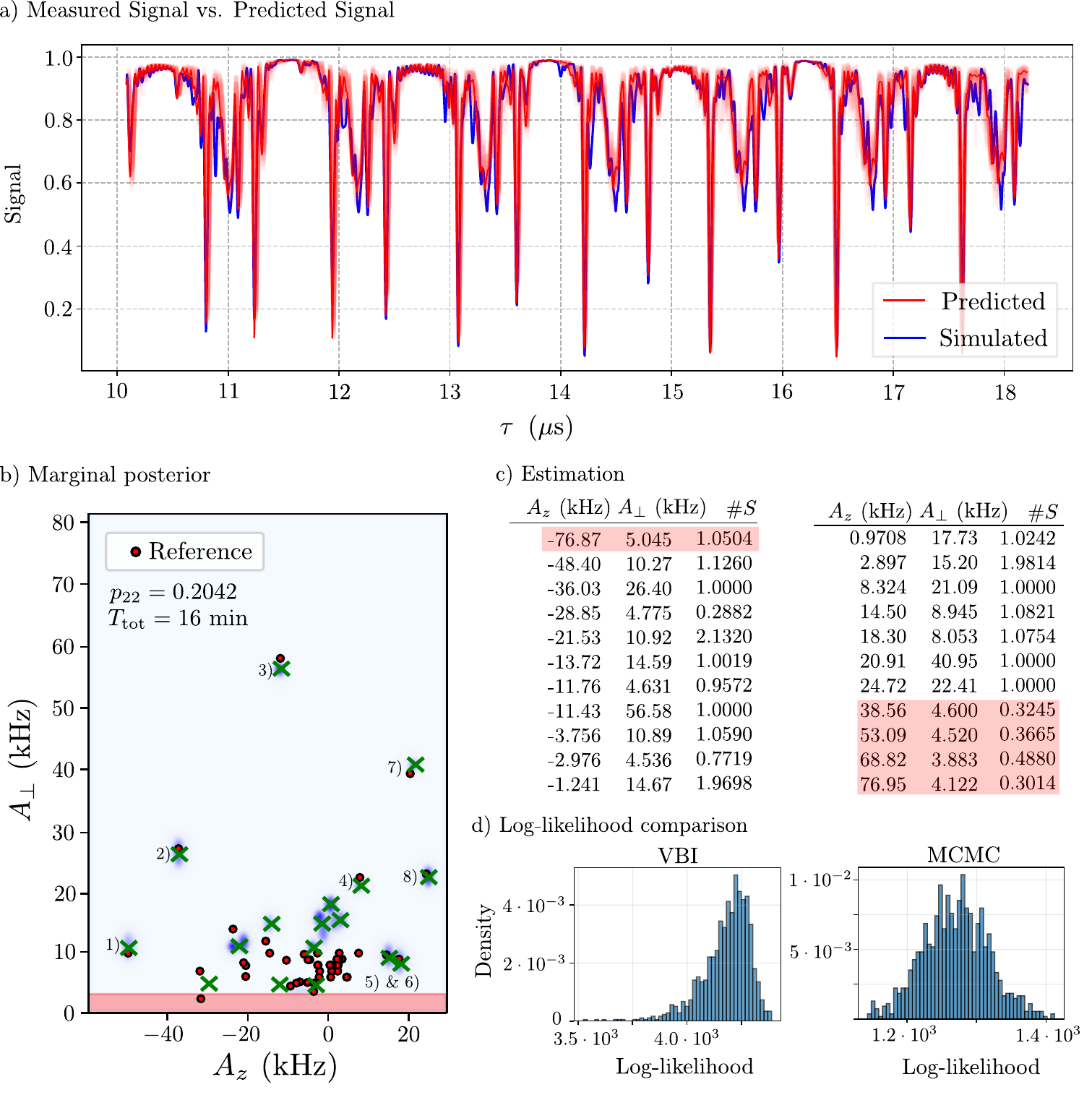}
    \caption{\justifying 
    \textbf{Benchmarking the algorithm on experimental data.} 
    \textbf{a)} Dynamical decoupling signal [Eq.~\ref{eq:nanonmr_model}] as a function of the inter-pulse delay $\tau$ for a subset of the data from M.~H.~Abobeih \textit{et al.}~\cite{abobeihOnesecondCoherenceSingle2018}, with $\tau$ roughly between 10 and 18~$\mu$s ($M=2048$ measurement points with $4$\,ns between each data point). The blue curve represents the experimental data, and the red curve shows the fit obtained using the VBI algorithm. The prediction uncertainty is indicated by the light red lines around the prediction, corresponding approximately to one standard deviation. 
    \textbf{b)} Marginal posterior distribution trained on the observed signal, conditioned on the most likely number of spins, identified by the algorithm as $n=22$. A total of $K=40$ spins are fit to model the weakly interacting spin bath and better reproduce the signal. The red dots represent the spins estimated by the deep-learning approach of~\cite{jungDeepLearningEnhanced2021} using the $N_{\pi}=32$ and $N_{\pi}=256$  complete datasets, which required approximately 4\,h of data acquisition. \textbf{c)} List of the hyperfine values and corresponding weights for the identified spins, divided into two tables according to the sign of $A_z$. Spins highlighted in red indicate false positives. Around the origin, more spins are not identified precisely. See the caption of Fig.~\ref{fig:simulated_512_meas} for interpretation of the marginal posterior. 
    \textbf{d)} Comparison of the log-likelihood distributions obtained with VBI and MCMC. We consider Metropolis–Hastings MCMC with an optimized Gaussian proposal, similar to \cite{poteshman_HighthroughputSpinbathCharacterization_2025}. We use a single chain, extracting $10^4$ samples and fixing the number of spins to $50$ (no reversible jump is implemented here).}
    \label{fig:experimental}
\end{figure*}

As illustrated in Fig.~\ref{fig:experimental}, we are able to successfully identify at least $8$ spins from the experimental data. All the other identified spins above threshold, and below threshold, contribute to the fit of the signal. As ground truth values are not available in the experiment, we report with red dots the hyperfine couplings estimated in previous work by K. Jung \textit{et al.}~\cite{jungDeepLearningEnhanced2021}. In~\cite{jungDeepLearningEnhanced2021} the authors identify $14$ spins using only the same $N_{\pi}=32$ DD dataset that we have processed, but extended to $\tau \in [6, 50] \,\mu s$. The training time of the posterior for the experimental data is $8$ min, to be compared with $8$\,h for the model training and $50$\,s of processing for each spin of the deep learning approach.

The spins not used in the explanation of the signal (i.e. with couplings close to zero) are pushed into the blind zone by regularization and pruned by the thresholding. The confidence that the environment comprises $22$ spins is $p_{22} = 0.2042$, the total signal acquisition time ($T_\text{tot}$) is $16$ minutes. 

An interesting question is how the performance of our VBI-based approach compares to MCMC, utilized for example in Ref.~\cite{poteshman_TransdimensionalHamiltonianModel_2025} in fitting the experimental data. To assess this, we have implemented a simple Metropolis-Hastings MCMC for a fixed number of parameters and without parallel tempering (different than the reversible-jump implementation, with parallel tempering described in Ref ~\cite{poteshman_TransdimensionalHamiltonianModel_2025}).

In Fig.~\ref{fig:experimental}(d), we report histograms of the log-likelihoods resulting from the applications of, respectively, VBI (left) and MCMC (right). VBI reaches higher log-likelihood values than our MCMC baseline, indicating a better fit within our model class. In addition, the estimation time reported in Table~\ref{table:training_time} scales more favorably with the number of data points for VBI than for Metropolis–Hastings MCMC in our implementation, making VBI more suitable for rapidly updating a posterior in the context of online Bayesian experimental design.

While MCMC is able to fit a number of spins, it appears to not be necessarily converging to the right mode of the posterior. On the other hand, VBI is probably underestimating uncertainties, and therefore it remains closer to the maximum likelihood estimator than MCMC. VBI also appears to lead to heavy tails in the log-likelihood histograms, which are absent in the MCMC histograms. A more accurate and complete comparison between VBI and MCMC, also involving reversible jumps, is out of the scope of this manuscript and is left to future work.

A natural question is how our approach compares to methods based on deep learning~\cite{jungDeepLearningEnhanced2021}, in terms of the number of spins successfully detected and the associated estimation errors in the limit of large datasets. We report a detailed performance analysis in Appendix~B, showing that the architecture utilized in this work is unable to match the results of~\cite{jungDeepLearningEnhanced2021}, even when increasing the number of layers in the normalizing flow. In particular, we observe that the training process slows down significantly as the ansatz complexity increases; numerical instabilities appear in the gradient descent already at $32$ layers, accompanied by prohibitive memory requirements. We conjecture that these challenges originate from the highly peaked structure of the likelihood when large datasets are used.

A comprehensive investigation of alternative architectures to optimize variational Bayesian inference for individual nuclear spin identification in the large-data regime is beyond the scope of this manuscript. However, our observation that increasing the number of autoregressive layers improves the fit quality indicates that capturing the non-Gaussian structure of the posterior peaks is essential. This, in turn, suggests that greater expressive power is required from the ansatz, for example through the use of a mixture of flows  or a diffusion-based posterior representation.

\subsection{Methods}
\label{subsec:methods}
The algorithm has been coded using Pyro~\cite{bingham_PyroDeepUniversal_2018, phan_ComposableEffectsFlexible_2019}, a universal probabilistic programming language (PPL)~\cite{baudartDeepProbabilisticProgramming2018} written in Python and supported by PyTorch on the backend. All simulations have been run on an NVIDIA A40 GPU with 48\,GB of dedicated VRAM, while the CPU simulation have been executed on an Apple M1. The training of the posterior ansatz has been executed with the ADAM optimizer and an exponentially decaying learning rate schedule, interpolating between an initial and final learning rates that were respectively set to $\ell(I=1) = 10^{-3}$ and  {$(I = 8196) = 10^{-5}$}.

\section{Discussion and outlook}
\label{sec:outlook}

This work presents a framework for parameter estimation and model selection for quantum experiments, based on variational Bayesian inference. We show that VBI enables the characterization of quantum systems with a number of parameters significantly beyond what is currently feasible with techniques such as particle filtering \cite{granadeRobustOnlineHamiltonian2012}. We expect this framework to be broadly applicable across many different physical platforms, to high-dimensional estimation problems related to quantum tomography \cite{huszarAdaptiveBayesianQuantum2012, wiebeEfficientBayesianPhase2016}, Hamiltonian learning \cite{granadeRobustOnlineHamiltonian2012, wiebeQuantumHamiltonianLearning2014, valenti_HamiltonianLearningQuantum_2019, gentileLearningModelsQuantum2021, valenti_ScalableHamiltonianLearning_2022}, the characterization of NISQ quantum circuits \cite{duffield_BayesianLearningParameterised_2023}, multi-parameter quantum sensing \cite{pezze_AdvancesMultiparameterQuantum_2025}, and others.

In addition to parameter estimation for high-dimensional systems, our framework includes the capability to select models within a class of nested models featuring different numbers of unknown parameters. This is a topical challenge of great practical importance in quantum science and technology, as it captures the physics of a known quantum system interacting with an environment where not all relevant degrees of freedom might be known. Examples of this are a spin qubit coupled to an environment comprising an unknown number of other electronic and nuclear spins \cite{taminiauDetectionControlIndividual2012, abobeihOnesecondCoherenceSingle2018, zopesThreeDimensionalNuclearSpin2018, budakianRoadmapNanoscaleMagnetic2024}, a superconducting qubit coupling to an unknown number of two-level fluctuators \cite{burnett_EvidenceInteractingTwolevel_2014, degraaf_TwolevelSystemsSuperconducting_2020}, the explainable identification of the processes governing the dynamics of an unknown quantum system \cite{wallace_LearningDynamicsMarkovian_2025, fioroniLearningAgentbasedApproach2025}, and generally multi-qubit systems featuring an unknown number of coherent coupling and decoherence processes (but limited within a known library of processes).

We have benchmarked our approach on a specific problem of high technological interest, i.e. the identification of individual nuclear spins from measurements on a single electron spin quantum sensor. This problem is sufficiently complex, in terms of number of free parameters and nonlinearity of the signal, to showcase the power of our framework in tackling the characterization of multi-dimensional quantum systems with unknown numbers of components. We have demonstrated, on simulated data and on real experimental data, how we can infer the hyperfine values for the spins that contribute to the observed signal, even when their number is not known in advance. Our protocol, parallelized on a GPU, can identify up to 16 spins in a few minutes, a timescale much shorter than that required for other Bayesian techniques such as Markov Chain Monte Carlo. 

While we have used nuclear spin detection as an example of the application of the technique, a direction of future work is to push the identification of nuclear spins to the limit of the largest number of spins and the smallest uncertainty on their hyperfine couplings and actual physical positions. This requires deploying the algorithm on larger datasets, with more values of the inter-pulse delay $\tau$, as here we have only used a portion of the data in M.~H.~Abobeih et al.~\cite{abobeihOnesecondCoherenceSingle2018}. For sufficiently large values of $\tau$, the dipolar coupling between nearby nuclear spins becomes important, requiring updating the Hamiltonian in Eq.~\ref{eq:nanonmr_hamiltonian} with the corresponding terms. The resulting Hamiltonian has analytical solutions for pairs of nuclear spins \cite{zhao_AtomicscaleMagnetometryDistant_2011, shiSensingAtomicscaleStructure2014, abobeihOnesecondCoherenceSingle2018, herbMultidimensionalSpectroscopyNuclear2024}, while numerical solutions can be found with the cluster expansion method for larger spin clusters \cite{mazeElectronSpinDecoherence2008, yangQuantumManybodyTheory2008, yang_QuantumManybodyTheory_2009, yang_QuantumManybodyTheory_2017, onizhuk_PyCCEPythonPackage_2021}. Furthermore, the learning will have to be integrated with different types of measurements, such as the DD-RF sequence \cite{bradleyTenQubitSolidStateSpin2019} interleaving dynamical decoupling with phase-controlled driving of nuclear spins, and correlated double-resonance sequences to map connected chains through the network of spins\cite{vandestolpeMapping50spinqubitNetwork2024}. To give an idea of the power of these techniques, and the complexity of the sensing and data processing procedures, these have recently enabled mapping a 50-nuclear-spin network comprising 1225 spin-spin interactions in the vicinity of an NV center \cite{vandestolpeMapping50spinqubitNetwork2024}. This is an important application as the measurement time is much longer than for the dynamical decoupling datasets presented here, leading to potentially big advantages. Additionally, not all couplings are required to resolve the structure, and adaptive protocols can therefore be extremely advantageous in deciding which ones to characterize based on information from the previous measurements.

Our framework is purely based on classical algorithms and, as with any classical algorithm processing quantum systems \cite{gebhartLearningQuantumSystems2023, krenn_ArtificialIntelligenceMachine_2023,dawid_MachineLearningQuantum_2025}, complexity scales exponentially with size. It is therefore most useful for a number of parameters which is too large for exact Bayesian inference but still sufficiently small for an efficient classical simulation of the probability outcomes that are required for applying VBI. Our techniques are thus best suited to problems where the quantum system of interest is simple enough to permit efficient evaluation of its dynamics, but where the number of parameters characterizing it is sufficiently large {to constitute} a high-dimensional parameter estimation problem. The example described in Section~\ref{sec:nanonmr_example} of this manuscript fits in this sweet spot.

It is also important to note that, while we train the full multi-dimensional posterior, for the clustering analysis that gives us an estimator for the couplings we do not leverage any of the correlations between the spins, discarding some information captured by the posterior for simplicity. In an adaptive scenario, the full posterior would be used to decide the next optimal measurements to perform, including correlations between the parameters and confidence levels.

Our parameter learning and model selection algorithm produces an approximate posterior for all the models and parameters for the quantum system, given a certain set of measured data. The algorithm exploits parallelization, leading to short computation times on the order of a few minutes with modern GPUs (see Table~\ref{table:training_time}). This fast-computed approximate posterior can be used to optimally choose subsequent measurements and probe state design in order to maximize the final precision of the estimation, in optimal Bayesian experimental design~\cite{rainforthModernBayesianExperimental2024}. There are different possibilities to develop heuristics for the optimization of subsequent measurements, ranging from information-theoretic quantities such as Fisher information or information gain, to deploying reinforcement learning \cite{reuer_RealizingDeepReinforcement_2023, belliardoModelawareReinforcementLearning2024, belliardoApplicationsModelawareReinforcement2024}. While information-theoretic quantities are typically expensive to compute, our framework can be extended to approximate them. For example, one could use the surrogate information gain:
\begin{equation}
    \text{SIG}(\vx) := \mathbb{E}_{\vy} \lbrace \text{Var}_{\vtheta \sim q_{\vlambda}(\vtheta)} \left[ p( \vy \mid \vx, \vtheta) \right] \rbrace ,
    \label{eq:surrogate_IG}
\end{equation}
which is easy to compute, since the variance is computed on the distribution $q_{\vlambda}(\vtheta)$, and $p(\vy|\vx, \vtheta)$ is the probability for the observation of the outcomes $\vy$ given the controls $\vx$. We could then sample the next measurements among those that feature a large SIG, realizing thereby optimal experimental design based on VBI~\cite{sarraDeepBayesianExperimental2023}.

The posterior training times (Table~\ref{table:training_time}) highlighted for the nanoscale magnetic resonance example in Section~\ref{sec:nanonmr_example} are suited for integration of adaptivity, with the goal of greatly reducing data acquisition time. As a reminder of the importance of smart data acquisition for this application, the total sensing time for a full dynamical decoupling trace with inter-pulse delay $\tau$ ranging from $10\,\mu$s to $300\,\mu$s and 4\,ns discretization \cite{abobeihOnesecondCoherenceSingle2018} is more than 8 days ($N_{\pi}=32$, $R=10^3$ repetitions). This is the bare sensing time, neglecting non-deterministic overheads due to the NV charge and spectral instabilities, the initialization of the NV's $^{14/15}$N nuclear spin, and any setup-related issues such as periodic optical re-alignment, calibration, or drifts. 

We envision adaptivity to be deployed by grouping data acquisition in batches. Training the posterior for a batch of 1024 measurements takes about 6 minutes (Table~\ref{table:training_time}). Data acquisition for 1024 values of the inter-pulse delay $\tau$ takes about 9 minutes in the range $\left[6, 10\right]\,\mu$s and more than one hour for $\tau \in \left[60, 64\right]\,\mu$s. The goal of adaptivity will be to utilize “cheap” information gained at shorter $\tau$ or smaller number $N_{\pi}$ of pulses to prioritize measurements with the largest information content for larger values of $\tau$ and $N_{\pi}$, where data acquisition is more time-consuming. Posterior training durations comparable to the shortest data acquisition batches offer a good investment for the expected large gains in avoiding long measurements with small information content. Our preliminary analysis \cite{varona-uriarte_ComputationallyTractableOffline_2025} suggests that the amount of DD data collected can be greatly reduced without sacrificing estimation performance. 

The data acquisition timescales described here refer to experiments performed at low temperature on NV centers deeper inside the diamond \cite{abobeihOnesecondCoherenceSingle2018}, where fast high-fidelity single-shot electron spin readout is available \cite{robledo_HighfidelityProjectiveReadout_2011}. Room-temperature experiments and/or NV centers close to the diamond surface often use averaged readout; while Bayesian inference is still feasible and advantageous \cite{dinaniBayesianEstimationQuantum2019}, the lack of single-shot readout leads to a significant slowdown of data acquisition, which makes the posterior training time completely negligible.

The operation of any technology requires dealing with imperfections, such as fabrication flaws, instabilities, and the impact of uncontrolled environmental factors. All these factors are often complicated to model and predict yet need to be included in the Bayesian inference process; otherwise, they will lead to incorrect estimation and sub-optimal (if not detrimental) feedback control loops. A promising avenue for future exploration is to integrate VBI with “gray-box’’ techniques \cite{_PracticalGreyboxProcess_2006, genois_QuantumTailoredMachineLearningCharacterization_2021, youssry_ExperimentalGrayboxQuantum_2024} which couple a physics-based model for the system with a data-driven model for the imperfections, which can be dependent on external parameters such as temperature, electric/magnetic fields, etc.

Another interesting open question is how to account for symmetries in the problem \cite{bilosScalableNormalizingFlows2021, rasulSetFlowPermutation2019, zwartsenbergConditionalPermutationInvariant2022, maronLearningSetsSymmetric2020, thiedeGeneralTheoryPermutation2020, zhouPermutationEquivariantNeural2023, tolooshamsEquiRegEquivarianceRegularized2025, kohlerEquivariantFlowsExact2020}, such as the permutational symmetries of spin labeling for the example described in Section~\ref{sec:nanonmr_example}. Assuming that the prior model presents the same symmetries, the exact posterior $P(\vtheta | \vx, \vy)$, calculated from {Bayes'} rule, inherits the symmetries of the model $p(\vtheta | \vx, \vy)$. However, the variational ansatz for the posterior distribution does not necessarily preserve these symmetries. For discrete symmetries, the posterior distribution exhibits multiple equivalent modes in the parameter space. In this case, the reverse KL-divergence used in variational inference (Eq.~\eqref{eq:elbo_definition}) is known to manifest a mode-seeking behavior, which theoretically leads to the posterior concentrating on a single mode. In practice, during training, the posterior probability density tends to concentrate in only a few of the multiple equivalent regions of the parameter space, neglecting the others. An interesting future direction is to explore the use of symmetry-preserving ansatzes, with the idea that by stopping the ansatz from “leaking’’ into other modes, its expressivity will be better exploited in the approximation of the posterior.

\section{Data availability}

All experimental and simulated data used in this work is available in the code repository. The experimental data is published in M. H. Abobeih \textit{et al.} \cite{abobeihOnesecondCoherenceSingle2018}.

\section{Code availability}
The open source library BrightCarbon can be found in the \textbf{\url{https://gitlab.com/federico.belliardo/brightcarbonsoftware}} repository and can be installed with pip following the instructions in the README file.

\section{Acknowledgments}
We thank D. Yudilevich, B. Varona-Uriarte and L. Sarra for useful discussions. We gratefully acknowledge the computational resources of the ``High Performance Computing'' facility at Heriot-Watt University. This work is funded by the Engineering and Physical Sciences Research Council (EP/S000550/1, EP/V053779/1, EP/Z533208/1, EP/Z533191/1) and the European Innovation Council (QuSPARC, grant agreement 101186889). This work is also supported by the project 23NRM04 NoQTeS, which has received funding from the European Partnership on Metrology, co-financed from the European Union’s Horizon Europe Research and Innovation Programme and by the Participating States. {Finally, this project has received funding from the European Research Council (ERC) under the European Union’s Horizon 2020 research and innovation programme (grant agreement No. 852410)}

\section{Author Contributions}
C. B., F. B., Y. A. conceived the project. F. B. developed the algorithm (with input by Y. A. and C.B.), carried out the simulations and analyzed the results (in collaboration with C. B., Y. A. and E. M. G.). M. H. A. and T. H. T. provided the experimental data. F.B. and C.B. led the writing of the manuscript, with contributions from all co-authors.

\appendix

\section{Intuitive interpretation of the experiment}
\label{sec:intuitive}
In this Appendix, we discuss intuitively the properties of the Hamiltonian in Eq.~\eqref{eq:nanonmr_hamiltonian}. We perform an estimation with $N_{\pi}$ fixed and $\tau \in (\tau_{\min}, \tau_{\max})$ uniformly distributed. In the limit of large magnetic field (i.e. $f_L \gg \lbrace A_{z, k} \rbrace$), the probability distribution in Eq.~\eqref{eq:nanonmr_model} is characterized by multiple resonances for each spin, at values of the inter-pulse delay $\tau$ \cite{taminiauDetectionControlIndividual2012} given by Eq. ~\ref{eq:tau_resonance}. Each resonance is Lorentzian, with width $\mathcal{O}\left( \frac{A_{\perp, k}}{f_L^2} \right)$ and depth $\mathcal{O}\left( \frac{N_\pi A_{\perp, k}}{f_L}\right)$.

From the requirement that the distance between the location of two dips is greater then the width of the resonances we can define a \textit{Rayleigh criterion} for being able to individually identify multiple spins in the environment. This tells us that we need to look at least at the resonances of order
\begin{equation}
    m_{\min} := \frac{\max A_\perp}{4 \min \left( \Delta A_{z} \right)} \; ,
\end{equation}
to distinguish every spin. Here, $\min \left( \Delta A_{z} \right)$ indicates the minimum distance between the hyperfine couplings of each pair of spins in the detection volume. For realistic numbers (e.g. $\max A_\perp = 80 \kHz$), resolving hyperfines with $\min \Delta A_{z} = 0.1 \kHz$ requires taking data for resonances up to $m_{\min} = \mathcal{O}(10^{2})$. This is a fairly time-consuming data acquisition, particularly as larger values of $m$ result in a larger interpulse delay $\tau$ and, consequently, longer data acquisition. A possible strategy to address this is to first take data on resonances with lower $m$, e.g. $m < \mathcal{O}(10)$, and then use adaptive Bayesian experimental design to select the specific inter-pulse delays $\tau$ corresponding to resonances in the $m_{\min} = \mathcal{O}(10^{2})$ range that give the most information in distinguishing the required spectral shifts, based on the information accumulated. The development of optimal adaptive schemes for nuclear spin identification is however beyond the scope of this manuscript, and will be the subject of future dedicated work. 

In order to detect all the resonances, the discretization in the inter-pulse delay ($\delta \tau$)  should be of the order of the width of the resonance, i.e. $A_{\perp} / {f_L}$. The $m$-th resonances fit within a window of size $1 / 4 f_L$, giving a number of discrete inter-pulse delay values of
\begin{equation}
    M_{\min} := \frac{1}{4 f_L} \left( \frac{\min{A_\perp}}{f_L^2} \right)^{-1} = \mathcal{O}\left( \frac{f_L}{4 \min{A_\perp}} \right) \; . 
    \label{eq:min_measurement}
\end{equation}
For $\min A_{\perp}  = 8 \kHz$ and $f_L = 0.43 \MHz$ we get $M_{\min} \simeq 85$, which is in the ball park of the number of different $\tau$ needed to start distinguishing the spins individually in our simulations. Additionally, for a dip to be detectable, the shot-noise must less than its depth. This sets a condition on the number of repetitions $R$ for each DD sequence, which reads
\begin{equation}
    \sqrt{\frac{p(1-p)}{R}} \le \frac{N_\pi \min A_\perp}{f_L} \; .
\end{equation}
where
\begin{equation}
    p := 1 - \frac{N_\pi \min A_\perp}{f_L} \; .
\end{equation}
and this equation is effectively just noise $<$ signal. For small dips the condition on $R_{\min}$ is
\begin{equation}
    R \ge R_{\min} := \frac{f_L}{N_{\pi} \min{A_\perp}} \; ,
\end{equation}
Combining all the previous considerations, the condition on the total measurement time for being able to distinguish all the spins is
\begin{equation}
    T_{\min} = \mathcal{O} \left( \frac{f_L \max{A_\perp}}{\min \Delta A_z (\min A_\perp)^2} \right) \; .
\end{equation}
For the numerical values reported above we get $T_{\min} \simeq 3$ s, which is at least two orders of magnitude smaller than the total measurement time in our simulations. This could be an indication that we could actually use far less time in nano-NMR experiments than currently done~\cite{varona-uriarte_ComputationallyTractableOffline_2025}, or maybe the prefactor of $T_{\min}$ is not of order one.

The discrepancy in the estimated time could be solved by assuming that there are additional sources of noise on the system, and therefore we need to increase $R_{\min}$, which for our range of parameters is only $\sim 4$, far from the used value of $R \simeq \mathcal{O} (10^3)$. Since we have assumed additional sources of noise in the analysis of this manuscript, the above analysis might indicate that the algorithm we propose works close to the theoretical limits in the individual identification of the spins.

To summarize, we have considered the large-field limit of the task of determining the hyperfine couplings for independent nuclear spins in the environment of a single electron spin, and we have established lower bounds on the time and measurements needed in the learning. In the limit $\omega_L \rightarrow \infty$, the task is equivalent to the problem of searching and identifying the resonances by scanning over an interval of times $\tau$s. This is similar to the setting of Allen \textit{et al.} \cite{allenQuantumComputingEnhanced2025}, suggesting that the estimation could be sped up by quantum algorithms, to achieve the Heisenberg-Grover limit.

\section{Extending the dataset and the expressivity of the ansatz}
\label{eq:more_data}
\begin{figure*}[!htbp]
    \centering
\includegraphics[width=\textwidth]{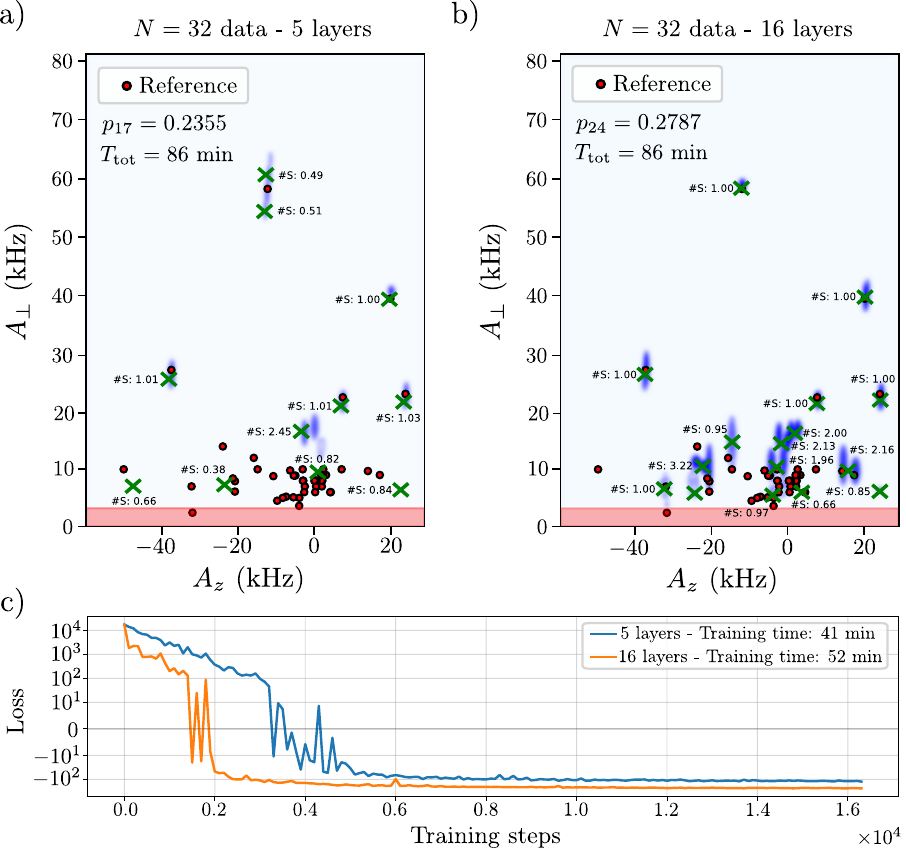}
    \caption{\justifying Result of the fit for the dataset with $N_\pi = 32$ for $\tau \in [6, 50] \, \mu s$. a) Restricted posterior obtained using a normalizing flow ansatz with $5$ autoregressive layers. b) Restricted posterior obtained with $16$ autoregressive layers. In both posteriors, the results of the clustering algorithm are superimposed; green crosses indicate the cluster centers, with the number of spins in each cluster labeled accordingly. c) Loss function, specifically the negative ELBO from Eq.~\eqref{eq:elbo}, as a function of the training step. The $16$-layer ansatz converges more rapidly to the asymptote and achieves a lower loss, despite requiring more computational time per training step.} 
    \label{fig:N32}
\end{figure*}

\begin{figure*}[!htbp]
    \centering
\includegraphics[width=\textwidth]{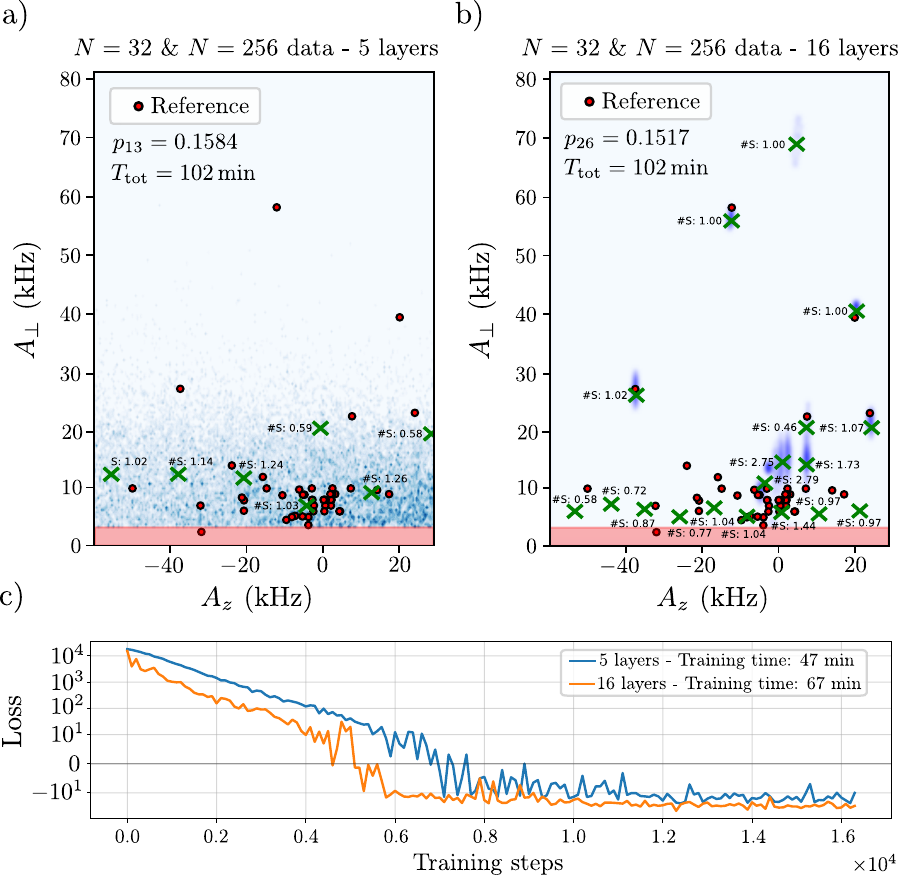}
    \caption{\justifying Reduced posteriors and training loss for the joint fit of datasets with $N_\pi = 32$ and $N_\pi = 256$ pulses. The description provided for Fig.~\ref{fig:N32} also applies here. Notably, while the final training loss values for the simple and complex architectures are comparable, the fit using fewer layers fails entirely to capture the characteristic features of the environment.}
    \label{fig:N32N256}
\end{figure*}

In this Appendix, we explore the performance of our architecture in the limit of a large dataset. We fit an extended dataset corresponding to $\tau \in [6, 50] \, \mu s$ for $N_\pi = 32$ and $\tau \in [18, 28] \, \mu s$ for $N_\pi = 256$. We first examine the fit for the $N_\pi = 32$ dataset alone (see Fig.~\ref{fig:N32}), which contains approximately $10^4$ points. We then jointly fit the data for $N_\pi = 32$ together with the $N_\pi = 256$ data (for a total of $\sim 1.25 \cdot 10^4$ points). Our results shows that, as the number of data points increases, the complexity of the variational ansatz becomes increasingly critical in determining the quality of the fit.

As in the main text, we employ a normalizing-flow ansatz constructed from either $5$ or $16$ autoregressive layers. In its base configuration, this architecture utilizes deep neural networks with $8$ hidden layers, each consisting of $256$ units.

In Fig.~\ref{fig:N32}, we compare the performance obtained on the full $N_\pi = 32$ dataset using two architectures: one with $5$ autoregressive layers and another with $16$. We also report the training loss (the negative ELBO from Eq.~\eqref{eq:elbo}). The reduced posterior indicates that the larger ansatz yields a substantially better fit to the data, a finding further supported by the lower asymptotic loss achieved during training. To ensure a fair comparison, all remaining hyperparameters were held fixed across training runs.

In Fig.~\ref{fig:N32N256}, we present the results of the combined fit using both the $N_\pi = 32$ and $N_\pi = 256$ datasets. For this more challenging setting, we employed $16$ autoregressive layers and increased the network depth from $8$ to $12$ hidden layers. Although the simpler ansatz fails completely, the more expressive architecture also produces unsatisfactory results, despite a substantial increase in training time from $47$\,min to $67$\,min.

We conjecture that increasing the ansatz complexity further may be necessary to obtain a satisfactory fit for datasets containing multiple pulse sequences; however, pursuing this direction lies beyond the scope of the present work. In particular, a more computationally efficient architecture is likely required.

In all reduced posteriors presented in Fig.~\ref{fig:N32} and Fig.~\ref{fig:experimental}, we include $47$ spins as a reference, following the identifications reported in~\cite{jungDeepLearningEnhanced2021}. We exclude a single spin with an abnormally large $A_z$ coupling that lies outside our analyzed window. The authors of~\cite{jungDeepLearningEnhanced2021} note that some redundancies may be present in their analysis, where a single spin is occasionally resolved as a pair of closely spaced spins, potentially reducing the effective number to $31$. For the $N_\pi = 32$ dataset, only $14$ of these spins have been confirmed through independent analyses in subsequent literature. By inspecting Fig.~\ref{fig:experimental} and Fig.~\ref{fig:N32}, and by relaxing the identification criteria to include clusters that are not perfectly centered on known spin locations, it is reasonable to argue that as many as $12$ spins can be resolved directly from the posterior, with additional spins originating from the surrounding bath.

When incorporating the $N_\pi = 256$ dataset, a significantly larger number of weakly coupled spins is identified, although only $7$ of them have been independently confirmed. In total, $21$ spins identified by the deep-learning method of~\cite{jungDeepLearningEnhanced2021} have received external confirmation. Finally, we observe in Fig.~\ref{fig:N32N256} that the $16$-layer ansatz fails to resolve the spins with the strongest couplings, instead fitting a large number of spins with comparatively smaller coupling strengths.

\bibliography{references}

\end{document}